\documentclass[pdflatex,sn-mathphys-num]{sn-jnl}

\usepackage{graphicx}%
\usepackage{multirow}%
\usepackage{amsmath,amssymb,amsfonts}%
\usepackage{amsthm}%
\usepackage{mathrsfs}%
\usepackage[title]{appendix}%
\usepackage{xcolor}%
\usepackage{textcomp}%
\usepackage{manyfoot}%
\usepackage{booktabs}%
\usepackage{algorithm}%
\usepackage{algorithmicx}%
\usepackage{algpseudocode}%
\usepackage{listings}%

\usepackage{mathtools}%
\usepackage{orcidlink}
\usepackage{setspace}
\usepackage{lmodern}
\usepackage{anyfontsize}
\usepackage{geometry}

\newgeometry{margin=3cm}

\begin{document}

\title{\centering
A likelihood-based Bayesian inference framework \\ for the calibration of and selection between \\ stochastic velocity-jump models}

\author*[1]{\fnm{Arianna} \sur{Ceccarelli}\orcidlink{0000-0002-9598-8845}}\email{ceccarelli@maths.ox.ac.uk}

\author[2]{\fnm{Alexander P.} \sur{Browning}\orcidlink{0000-0002-8753-1538}}

\author[3,4]{\fnm{Tai}\sur{Chaiamarit}}

\author[3,5]{\fnm{Ilan}\sur{Davis}\orcidlink{0000-0002-5385-3053}}

\author[1]{\fnm{Ruth E.} \sur{Baker}\orcidlink{0000-0002-6304-9333}}

\affil[1]{\centering
Mathematical Institute, University of Oxford, UK}
\affil[2]{\centering
School of Mathematics and Statistics, University of Melbourne, Australia}
\affil[3]{Department of Biochemistry, University of Oxford, UK}
\affil[4]{Department of Physiology, Faculty of Science, Mahidol University, Thailand}
\affil[5]{School of Molecular Biosciences, University of Glasgow, UK}

\abstract{Advances in experimental techniques allow the collection of high-resolution spatio-temporal data that track individual motile entities. These tracking data can be used to calibrate mathematical models describing the motility of individual entities. The challenges in calibrating models for single-agent motion derive from the intrinsic characteristics of experimental data, collected at discrete time steps and with measurement noise. We consider motion of individual agents that can be described by velocity-jump models in one spatial dimension. These agents transition between a network of \textit{n} states, in which each state is associated with a fixed velocity and fixed rates of switching to every other state. Exploiting approximate solutions to the resultant stochastic process, we develop a Bayesian inference framework to calibrate these models to discrete-time noisy data. We first demonstrate that the framework can be used to effectively recover the model parameters of data simulated from two-state and three-state models. Finally, we explore the question of model selection first using simulated data and then using experimental data tracking mRNA transport inside \textit{Drosophila} neurons. Overall, our results demonstrate that the framework is effective and efficient in calibrating and selecting between velocity-jump models and it can be applied to a range of motion processes.}


\maketitle

\section{Introduction}

Mathematical modelling is extensively used to characterize motion in biological systems, for example, to describe the motility of RNA~\citep{miles2025incorporating, miles2024inferring, harrison2019testing, harrison2018impact, ciocanel2018modeling, ciocanel2017analysis} and of molecular motors~\citep{han2024semi,hughes2012kinesins,hughes2011matrix,clancy2011universal,kutys2010monte}, the transport of intercellular cargoes along axons~\citep{cho2020fast,xue2017recent, bressloff2013stochastic, popovic2011stochastic, blum1988transport, blum1985model}, bacterial chemotaxis~\citep{guseva2025advantages,salek2019bacterial,rosser2014modelling, rosser2013novel, erban2004individual, berg1972chemotaxis} and the movement or migration of cells and animal species~\citep{mori2025optimal, pike2023simulating, powalla2022numerical, patel2018unique, jones2015inference, taylor2015birds, preisler2004modeling, medvinsky2002spatiotemporal, bovet1988spatial, kareiva1983analyzing}. Recent advances in experimental techniques allow the collection of data that track the location of individual agents with high spatio-temporal resolution. This kind of data is increasing in prevalence, motivating the development of models and frameworks to calibrate them, which can be used to describe and predict the behaviour of the motion process considered.

Single-agent motion is often modelled as a continuous stochastic process, while tracking data consists of a series of images collected with a fixed time frequency, from which agent locations are extracted with some intrinsic experimental noise (Figure~\ref{Fig_1}\textbf{A}). These inherent constraints of experimental data should not be neglected in the calibration of stochastic motion models. The study of these models and their calibration has been a long-standing area of mathematical research~\citep{liptser2013statisticsI, liptser2013statisticsII, doucet2001sequential}. A number of frameworks have been provided to estimate asymptotic or mean quantities of interest such as velocities and diffusivities~\citep{han2024semi,knoops2018motion,taylor2015birds}, or to estimate model parameters using data~\citep{harrison2018impact,rosser2013novel}. Existing frameworks to calibrate stochastic models to tracking data are typically based on likelihood-free approaches, such as particle filtering pseudomarginal methods~\citep{simpson2022reliable, warne2020practical, king2016statistical, andrieu2010particle}. Indeed, explicitly computing the likelihood of these continuous models subject to discrete-time noisy measurements is often not possible. Hence, there is a lack of likelihood-based inference methods suitable to calibrate these models. In this work, we tackle this issue by providing a framework suitable to estimate the parameters of stochastic velocity-jump models in one spatial dimension.

Velocity-jump models are suitable to describe motion characterised by the alternation between a number of constant velocities or pauses. For example, intracellular cargoes, which may comprise of mRNA, proteins, other molecules or organelles, are transported along the axons of the neurons by molecular motors that move processively in small steps on cylindrical cytoskeletal structures known as microtubules~\citep{encalada2014biophysical}. Microtubules have an intrinsic polarity and, along the axons, are often organised with their “minus ends” pointing towards the cell nucleus and their “plus ends” pointing to the periphery of the cell. There are two major classes of microtubule-dependent molecular motors, kinesins that predominantly carry cargoes in anterograde direction (towards the “plus ends” of microtubules at the axon terminal), and dyneins that predominantly transport them in retrograde direction (towards the “minus ends” of microtubules at the cell body)~\citep{gennerich2009walking}. Given the alternation of anterograde and retrograde movements along the same axis and pauses, axonal transport can be characterised by bi-directional, one-dimensional velocity-jump models. These models can capture the velocities at which the cargoes are transported in anterograde and retrograde directions, corresponding to the velocities in the forward and backward states, and the rates at which these changes in velocity occur. Furthermore, it is unclear whether the switching of direction requires a pause or can happen instantaneously.

To capture the behaviour of individual agents, we consider stochastic velocity-jump models in which the agent motion comprises a series of deterministic movements at constant velocity, separated by instantaneous reorientations, or jumps, during which the velocity may change~\citep{othmer1988models} (Figure~\ref{Fig_1}\textbf{A}). The agent motion is fully determined by the evolution of the internal hidden state of the agent, which evolves as a continuous-time Markov chain, and each state is characterised by a fixed velocity and fixed rates of switching to each other state. The difficulties in model calibration arise from the experimental limitation that the agent location is collected at discrete time steps and with measurement noise. This does not allow for determination of the precise state-switching times and, therefore, the direct computation of the model parameters. To address this, we construct a likelihood from an approximate probabilistic solution to the model derived in our previous work~\cite{ceccarelli2025} that accounts for noisy discrete-time tracking data. We use this likelihood approximation to build a Bayesian inference framework that allows for calibration of and selection between models using model selection criteria. In particular, we use the framework to tackle the question of selecting between models characterised by a different network of states or a different number of states.

In Section~\ref{Sec:Methods}, we present the methods used throughout the manuscript. First, we introduce the general $n$-state model and give an overview of the likelihood approximation and of the Bayesian inference framework based on a Metropolis-Hastings method~\citep{hastings1970monte,metropolis1953equation}. In Section~\ref{Subsec:Results_two-state}, we apply the framework to calibrate a simple two-state model capturing a forward-and-backward motion with two fixed velocities. Then, in Section~\ref{Subsec:Results_three-state}, we apply the framework to three-state models, in which the additional state represents a stationary phase in the agent motion. We also use the framework to select between models with different networks in Section~\ref{Subsec:Results_network_selection}, and with a different number of states in Section~\ref{Subsec:Results_model_selection}. In Section~\ref{Subsec:Results_experimental_data}, we apply the framework to calibrate two-state and three-state models to experimental tracking data and select between them to answer biological questions on the process of mRNA transport along \textit{Drosophila} axons, such as whether the switching between motion in opposite directions is direct or involves an intermediate stationary phase. Finally, in Section~\ref{Sec:Discussion_and_conclusions}, we discuss the main contributions of this work, emphasising that the approximate likelihood employed in our framework offers both reliability and reduced runtime compared to existing approaches, and we outline potential future directions.

\section{Methods}\label{Sec:Methods}

In this section, we introduce the $n$-state model and the approximation of the likelihood used, corresponding to the probability distribution function (PDF) of measuring discrete-time noisy data, explicitly computed in our previous work~\cite{ceccarelli2025}. Moreover, we develop a Bayesian inference framework based on a Markov chain Monte Carlo (MCMC) algorithm suitable to calibrate stochastic velocity-jump models in one spatial dimension to data. Finally, we introduce the criteria used to select between models.

\subsection{The motion model: a general stochastic \textit{n}-state velocity-jump model}\label{Subsec:Model}

We consider models for single-agent motion characterised by a fixed number of hidden states, $n$. Each state $s=1,\ldots,n$ is associated with a fixed velocity $v_s$, an exponential switching with fixed rate $\lambda_s$, and fixed probabilities to switch to any other state $u=1,\ldots,n$, $u\ne s$, $p_{su}$. These probabilities give the network probability matrix $P=[p_{su}]$ with zero diagonal entries. In line with the characteristics of experimental data, we assume that the agent location is only measured at fixed time intervals $\Delta t$ and with measurement noise, which we assume to be normally distributed with unknown standard deviation $\sigma>0$.

\subsection{The likelihood formulation}\label{Subsec:Likelihood}

\begin{figure*}[!ht]
    \centering
    \includegraphics[width=1\textwidth]{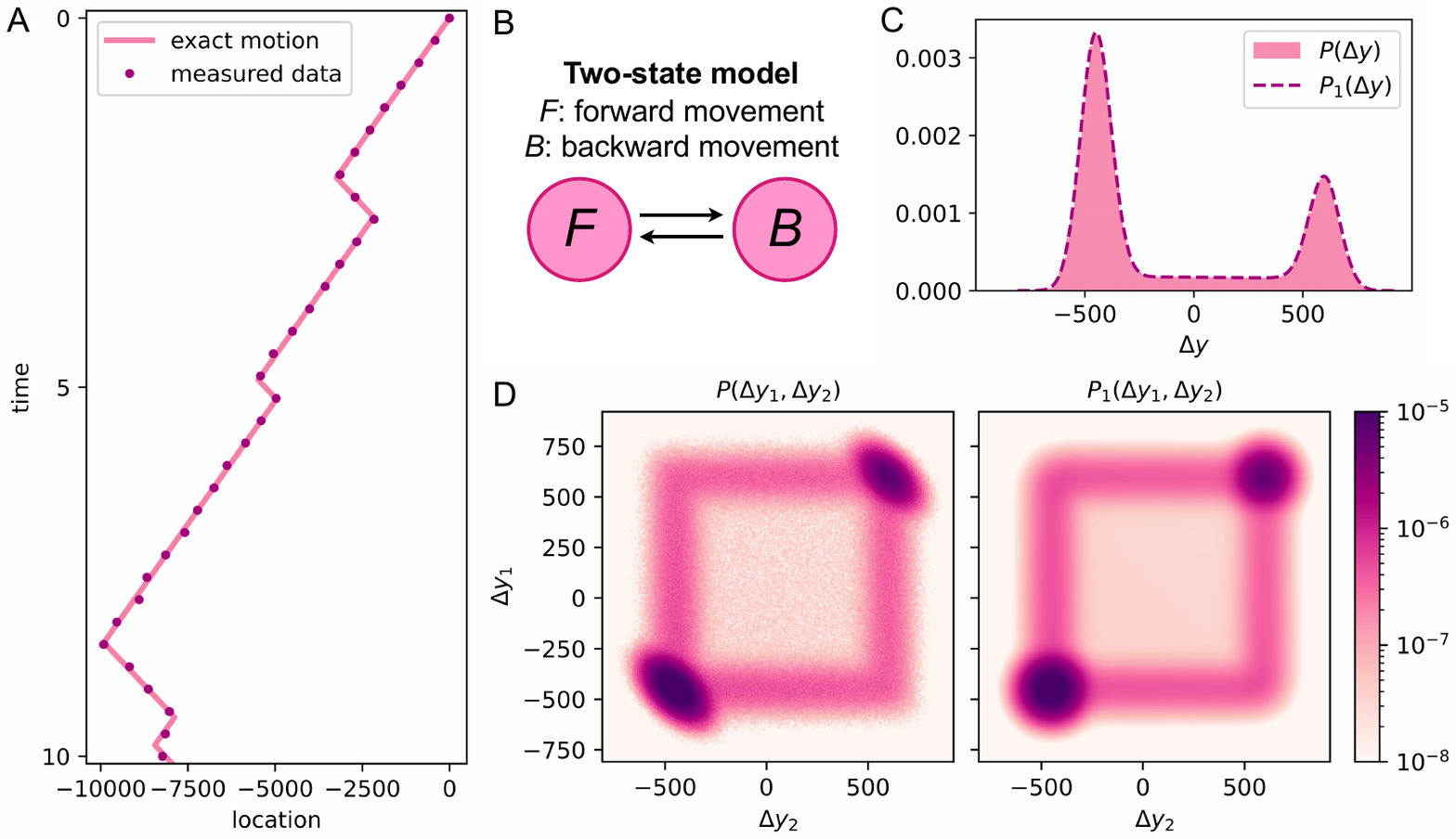}
    \caption{\textbf{Example of a two-state model}. \textbf{A} shows a simulation of the agent motion generated using the two-state model specified in \textbf{B}. \textbf{C} compares the empirical distribution for a single location increment $P(\Delta y)$, generated using simulated data (pink histogram) to the approximate distribution computed using the up-to-one-switch approximation for the PDF of measuring a single location increment, $P_1(\Delta y)$ (dashed line). \textbf{D} compares the track distribution for two subsequent location increments obtained using simulated data (left) and the approximation for the joint track distribution for two subsequent location increments $\Delta y_1$ and $\Delta y_2$, $P_1(\Delta y_1,\Delta y_2)$ (right).}
    \label{Fig_1}
\end{figure*}

We define a data track as a set of subsequently measured locations $[y_0,y_1, \ldots, y_N]$, (Figure~\ref{Fig_1}\textbf{A}). From a data track, we define a set of $N$ subsequent location increments as $\boldsymbol{\Delta y}_N:=[\Delta y_1, \Delta y_2, \ldots, \Delta y_N]$, where $\Delta y_i :=y_i - y_{i-1}$, for $i=1,2,\ldots,N$.

In our previous work~\cite{ceccarelli2025} we computed approximations for the PDF of measuring a single location increment $\Delta y$ by making the assumption that the agent switches state at most once between subsequent observations, defined as an up-to-one switch approximation for the model solution. Moreover, we also computed the up-to-one switch approximation for the probability of measuring a set of subsequent location increments $\boldsymbol{\Delta y}_N=[\Delta y_1, \Delta y_2, \ldots, \Delta y_N]$ obtained from a track. This approximation takes into account the correlation of subsequent location increments, due to the likely persistence in the same state.

Figure~\ref{Fig_1}\textbf{C} shows the up-to-one-switch PDF approximation for a single location increment $P_1(\Delta y)$, and compares it to the empirical PDF obtained using simulated data from an example two-state model. Figure~\ref{Fig_1}\textbf{D} shows the up-to-one-switch PDF approximation for two subsequent location increments $P_1(\Delta y_1, \Delta y_2)$, and compares it to the empirical PDF obtained using simulated data from the same two-state model. The methods to obtain these approximations and their accuracy are discussed in depth in~\cite{ceccarelli2025}. 

Across this work, we estimate a parameter set, denoted $\boldsymbol{\theta}$, which includes the velocities $v_s$, the natural logarithm of the switching rates $\log(\lambda_s)$, the probabilities required to define the network matrix $p_{su}$, and the noise standard deviation $\sigma$. We assume the time between measurements is a fixed, known quantity $\Delta t$. We refer to the \textit{track likelihood} associated with a parameter set $\boldsymbol{\theta}$, given the set of $N$ subsequent location increments obtained from a track $\boldsymbol{\Delta y}_N$, as $P_1(\boldsymbol{\Delta y}_N|\boldsymbol{\theta})$, which denotes the up-to-one-switch approximation for the joint PDF of a set of subsequent location increments obtained from a data track, $\boldsymbol{\Delta y}_N$, given the parameter set $\boldsymbol{\theta}$. In contrast, the \textit{marginal likelihood} is simply a product of the up-to-one-switch approximations of the PDF for each location increment,
$$P_1^{\mathcal{M}}(\boldsymbol{\Delta y}_N|\boldsymbol{\theta})=\prod_{i=1}^N P_1(\Delta y_i),$$
and, therefore, it does not take into account the correlation between data points. The marginal likelihood has the advantage of requiring considerably less runtime to compute compared to the track likelihood, as the correlations between subsequent location increments are not taken into account.

\subsection{Bayesian inference framework with Metropolis-Hastings MCMC}\label{Subsec:MCMC}

In this section, we present a Bayesian inference framework to estimate the parameters of a general $n$-state velocity-jump model for one-dimensional motion. The framework is suitable for calibrating a model using either a set of subsequent location increments obtained from a data track, multiple sets of subsequent location increments obtained from distinct tracks, or independent location increments obtained from distinct tracks.

The posterior distribution of the estimated parameter set $\boldsymbol{\theta}$ given a dataset $\mathcal{D}$ using the likelihood $\mathcal{L}$ is
\begin{equation}\label{Eq:posterior}
p(\boldsymbol{\theta}|\mathcal{D})\propto \mathcal{L}(\boldsymbol{\theta}|\mathcal{D})p(\boldsymbol{\theta}),
\end{equation}
where $p(\boldsymbol{\theta})$ is the prior distribution, representing the prior knowledge of the parameter set. In this work, we use approximations of the likelihood; hence we provide an approximation for the true posterior distribution. We estimate the switching rates on a logarithmic scale, and we choose uniform priors for all parameters, specified in Supplementary Information Section~S2 for every case. 

We sample the posterior distribution for the model parameters by running four chains using a Metropolis-Hastings algorithm~\cite{metropolis1953equation,hastings1970monte} for the MCMC method. Each chain starts at a parameter set $\boldsymbol{\theta}_0$ sampled from the specified uniform prior. We define a symmetric proposal density $q=q(\boldsymbol{\theta}, \boldsymbol{\Sigma})$, where $\boldsymbol{\Sigma}$ denotes the $k\times k$ covariance matrix of the set of $k$ estimated parameters, adaptively computed during the burnin phase (see~\cite{rosenthal2011optimal}). The algorithm, for $\hat n$ iterations, returns the estimated parameter posterior $\boldsymbol{\Theta}$ and the log-likelihood vector $\log\boldsymbol{\pi}$.

\begin{algorithm}\label{MCMC}
\caption{Metropolis-Hastings MCMC}
\begin{algorithmic}[1]
\State Given $q, \boldsymbol{\Sigma}, \boldsymbol{\theta}_0, \hat n, \mathcal{D}$
\State Set $k$ as the size of $\boldsymbol{\theta}_0$
\State Set $\boldsymbol{\theta} = \boldsymbol{\theta}_0$
\State Initialise $\boldsymbol{\Theta}$ as an $\hat n\times k$ array
\State Initialise $\log\boldsymbol{\pi}$ as a length $\hat n$ array
\State Compute $\log\pi = \log \mathcal{L}(\boldsymbol{\theta}|\mathcal{D})$
\State Save $\boldsymbol{\Theta}_{1,:}=\boldsymbol{\theta}$ and $\log\boldsymbol{\pi}_{1}=\log\pi$
\For {$j=2,3,\ldots,\hat n$}
    \State Sample a proposal parameter set $\boldsymbol{\theta}^* = q(\boldsymbol{\theta}, \boldsymbol{\Sigma})$
    \State Compute $\log\pi^* = \log \mathcal{L}(\boldsymbol{\theta}^*|\mathcal{D})$
    \State Sample the log acceptance probability
    $\log\alpha = \min(0, \log\pi^* - \log\pi)$
    \State Accept the transition with probability $\log\alpha$ 
    \If {$\log\alpha<R\sim U(0,1)$}
    \State Reassign $\boldsymbol{\theta} = \boldsymbol{\theta}^*$ and $\log\pi=\log\pi^*$
    \EndIf
    \State Save $\boldsymbol{\Theta}_{j,:}=\boldsymbol{\theta}$ and $\log\boldsymbol{\pi}_{j}=\log\pi$
\EndFor
\end{algorithmic}
\end{algorithm}

We implement the Metropolis-Hastings MCMC algorithm following~\cite{warne2020practical} and run four chains for each simulated dataset, first running the pilot chains to adaptively estimate the covariance matrix, then running $\hat n=10000$ MCMC iterations for each chain, which allow for convergence of the estimated posterior for each parameter according to the scale reduction factor ($\hat R <1.1$ for each parameter; see~\cite{gelman2013bayesian}). More details on the algorithm used are given in the Supplementary Information Section~S1.

\subsection{Model selection criteria}\label{Subsec:Criteria}

In Sections~\ref{Subsec:Results_network_selection} and \ref{Subsec:Results_model_selection} we use a criterion-based approach to select between models with different networks or numbers of states using in-silico data. The framework is used to sample model parameters from the posterior distribution independently for each model proposed and the results are compared using two selection criteria. Given a set of candidate models, the model chosen according to a criterion is the one with the lowest criterion value.

\subsubsection{Akaike information criterion}\label{Subsubsec:AIC}

The Akaike information criterion (AIC) is defined as
\begin{equation}\label{Eq:AIC}
    \mathrm {AIC} = 2k-2 \log{\hat {\mathcal{L}}},
\end{equation}
where $k$ is the number of estimated parameters and $\log{\hat {\mathcal{L}}}$ is the natural logarithm of the maximum likelihood for the model considered and is equivalent to $\max (\log \mathcal{L})$, obtained as the maximum across the approximations of the log-likelihood computed for the parameter sets in the posterior distribution. We note that the penalty term $2k$ ensures that, for small differences of $\log{\hat {\mathcal{L}}}$ between model A and model B (less than $|k_A-k_B|$), the model with the lowest number of parameters is chosen.

\subsubsection{Bayesian information criterion}\label{Subsubsec:BIC}

The Bayesian information criterion (BIC) is defined as
\begin{equation}\label{Eq:BIC}
    \mathrm {BIC} = k\log N-2 \log{\hat {\mathcal{L}}},
\end{equation}
where $k$ is the number of estimated parameters and $\log{\hat {\mathcal{L}}}$ is the natural logarithm of the maximum likelihood for the model considered and is equivalent to $\max (\log \mathcal{L})$. Again, in our case, up-to-one-switch approximations of the likelihood are used. In contrast with the AIC, the BIC also takes into account the number of data points used, $N$. In particular, when calibrating a dataset of $N$ location increments, the penalty term $k\log N$ ensures that, for small differences between $\log{\hat {\mathcal{L}}}$ for a model A and a model B (less than $|k_A-k_B|(\log N)/2$), the model with the lowest number of parameters is chosen.

\section{Results}\label{Sec:Results}

In this section, we present the results obtained using the framework to calibrate some example models to in-silico data generated using models of the form specified in Section~\ref{Subsec:Model}. For simplicity, we calibrate non-dimensional models and we fix the data collection frequency to $\Delta t=0.3$ in all cases.

\subsection{Parameter recovery is possible for the two-state model}\label{Subsec:Results_two-state}

\begin{figure*}[!ht]
    \centering
    \includegraphics[width=1\textwidth]{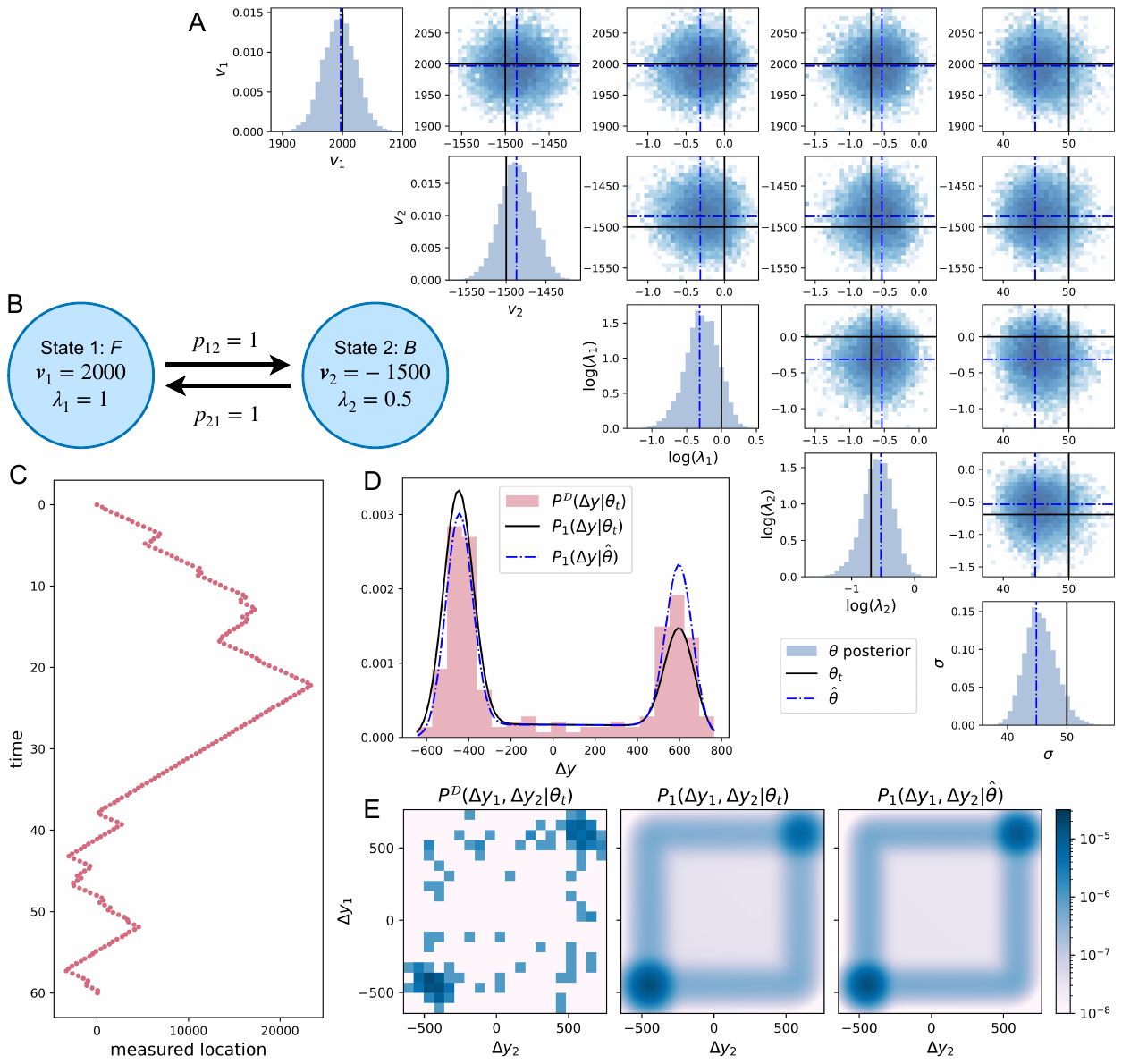}
    \caption{\textbf{Calibration of a two-state model to in-silico data}. \textbf{A} shows the univariate and bivariate distributions of the posterior of the estimated parameter set $\boldsymbol{\theta}=[v_1,v_2,\log(\lambda_1), \log(\lambda_2),\sigma]$. The black lines indicate the true parameter set, $\boldsymbol{\theta}_t$, and the blue dash-dotted lines indicate the estimated parameter set that gives the maximum likelihood $\boldsymbol{\hat\theta}$. \textbf{B} specifies the true parameters of the two-state model used to generate the agent motion. \textbf{C} shows the data track used in the calibration, collected with measurement noise $\sigma=50$ for a time interval of $T=60$, with fixed time between measurements $\Delta t= 0.3$, giving $N=200$ subsequent location increments. \textbf{D} shows the data distribution $P^\mathcal{D}$ (histogram in pink) compared to the likelihood $P_1$ at the true parameter set $\boldsymbol{\theta}_t$ (black line) and at the best parameter set $\boldsymbol{\hat\theta}$ (blue dash-dotted line) for a single location increment $\Delta y$. \textbf{E} compares the data distribution $P^\mathcal{D}$ for two subsequent location increments $[\Delta y_1,\Delta y_2]$ to the likelihood $P_1$ at the true parameter set $\boldsymbol{\theta}_t$ and at the best parameter set $\boldsymbol{\hat\theta}$.}
    \label{Fig_2}
\end{figure*}

We begin with the analysis of a synthetic dataset from which two states are evident (corresponding to the two peaks in the pink histogram in Figure~\ref{Fig_2}\textbf{D}). In Figure~\ref{Fig_2} we apply the framework using the track likelihood $P_1$ to the two-state model in Figure~\ref{Fig_2}\textbf{B}, which consists of a forward state and a backward state. The estimated parameter set is $\boldsymbol{\theta}:=[v_1, v_2, \log(\lambda_1), \log(\lambda_2), \sigma]$, and the true parameter set used to generate the data is $\boldsymbol{\theta}_t:=[2000,-1500,\log(1),\log(0.5),50]$. We note that we take the natural logarithm of the switching rates $\lambda$ to ensure equal emphasis is placed on these rates across different scales~\citep{raue2013lessons}. In this parameter regime, we show that the framework provided can be used to recover the true model parameters with significantly reduced variability in the posterior compared to the prior (Figure~\ref{Fig_2}\textbf{A} and Supplementary Information Figure~S1).

More generally, we highlight that the framework can be used to recover the parameters of a two-state model provided that state switches occur sufficiently infrequently compared to the data collection frequency (Supplementary Information Figure~S4) and the parameter posterior uncertainty is reduced as the number of data points is increased (Supplementary Information Figure~S5). The univariate and multivariate posterior distributions of the estimated parameter set $\boldsymbol{\theta}$ and the best parameter set denoted by $\boldsymbol{\hat \theta}$, that gives the maximum likelihood, tend to be close to the true parameter set $\boldsymbol{\theta}_t$, as shown in Figure~\ref{Fig_2}\textbf{A}. The uncertainty in parameter estimation is small compared to the priors (Supplementary Information Figure~S1). In Figure~\ref{Fig_2}\textbf{D}-\textbf{E} we also compare the approximation $P_1$ at the at the true parameter set $\boldsymbol{\theta}_t$ and at the estimated parameter set $\boldsymbol{\hat\theta}$ to the data distribution $P^{\mathcal{D}}$ than $P_1$.

We also calibrate a two-state model using the marginal likelihood $P_1^{\mathcal{M}}$, which is a simplified likelihood that assumes that the set of location increments provided are independent of one another, and it produces similar results to the track likelihood $P_1$ (see Supplementary Information Section~S2.1). The main advantage of the marginal likelihood over the track likelihood is the significantly lower runtime (Supplementary Information Figure~S5\textbf{F}). In particular, for the dataset in Figure~\ref{Fig_2}\textbf{C} of $N=200$ data points the runtime is reduced by three orders of magnitude. More generally, the runtime using the track likelihood scales linearly with the number of data points $N$, while it is approximately constant when using the marginal likelihood (Supplementary Information Figure~S5\textbf{F}).

\subsection{Models of increased complexity require more data for parameter identification}\label{Subsec:Results_three-state}

\begin{figure*}[!ht]
    \centering
    \includegraphics[width=0.89\textwidth]{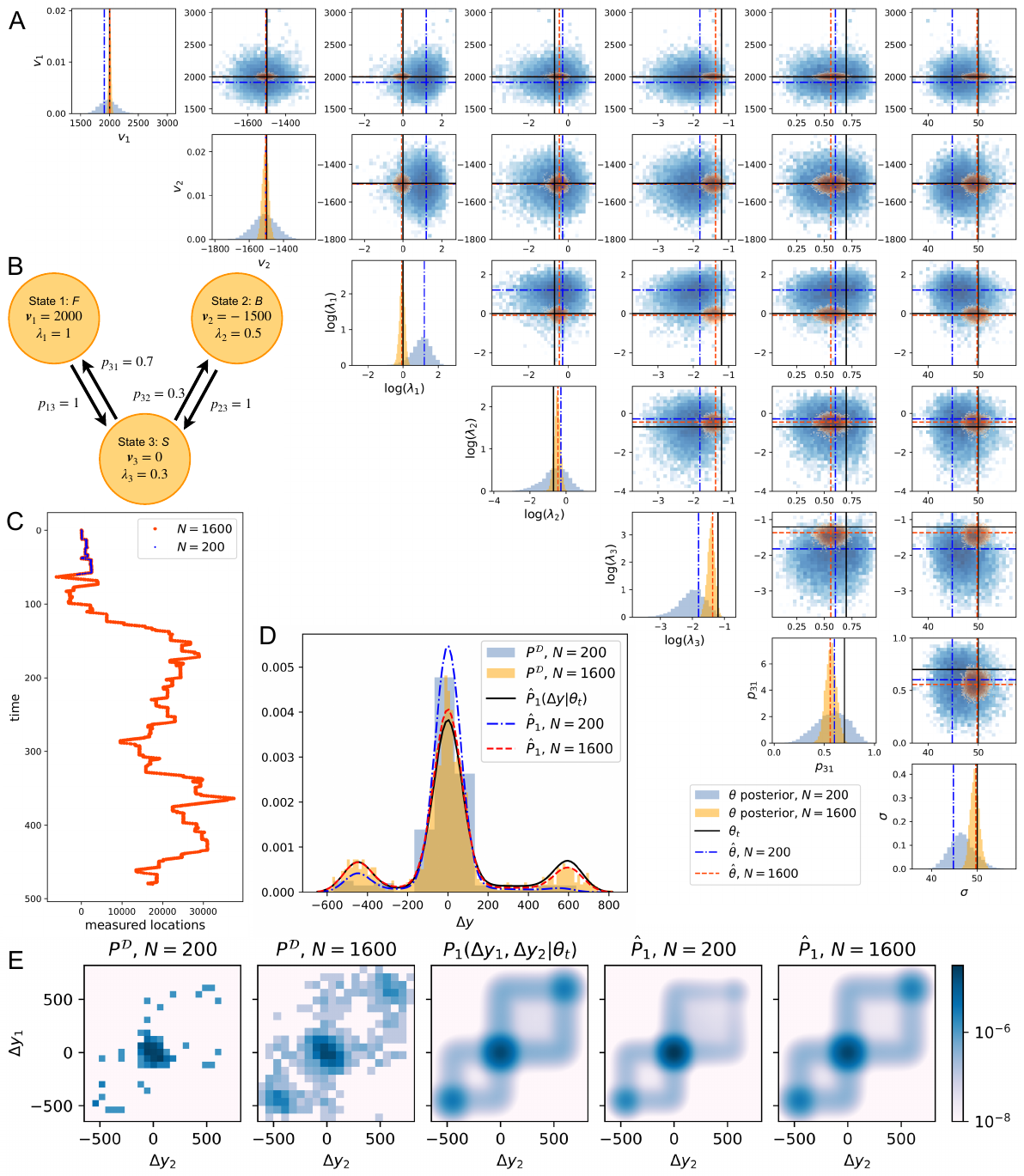}
    \caption{\textbf{Calibration of a three-state model given varying numbers of data points $N$}. \textbf{A} compares the univariate and bivariate distributions of the posterior of the estimated parameter set using a short ($N=200$, blue) and a long ($N=1600$, orange) track. The black lines in \textbf{A} indicate the true parameter set $\boldsymbol{\theta}_t$, and the orange and blue dashed lines indicate the parameter set that gives the maximum likelihood $\boldsymbol{\hat\theta}$ for $N=200$ and $N=1600$, respectively. The parameter set estimated is $\boldsymbol{\theta}=[v_1,v_2,\log(\lambda_1), \log(\lambda_2),\log(\lambda_3),p_{31},\sigma]$, while the velocity of the stationary state is fixed to zero ($v_3=0$), and we assume no direct switching between the forward and the backward states ($p_{12}=p_{21}=0$). The black lines indicate the true parameters $\boldsymbol{\theta}_t$ specified in \textbf{B}, and the dashed lines indicate the best parameter set $\boldsymbol{\hat\theta}$. Supplementary Information Figure~S6 shows these posteriors in comparison to their priors. \textbf{B} specifies the parameters of the three-state model used to generate the agent motion. \textbf{C} shows the two data tracks used, collected with measurement noise $\sigma=50$ with fixed time between measurements $\Delta t= 0.3$, for two time intervals of $T=60$ and $T=480$, that give $N=200$ and $N=1600$ subsequent location increments, respectively. \textbf{D} compares the data distributions $P^\mathcal{D}$ for a single location increment $\Delta y$ (histogram in blue for $N=200$ and in orange for $N=1600$) to the likelihood $P_1$ for a single location increment at the true parameter set $\boldsymbol{\theta}_t$ (black line) and at the best parameter sets estimated $\boldsymbol{\hat\theta}$ (blue dash-dotted and orange dashed lines). \textbf{E} compares the data distribution $P^\mathcal{D}$ for a set of two subsequent location increments $[\Delta y_1,\Delta y_2]$ to the likelihood $P_1$ at the true parameter set $\boldsymbol{\theta}_t$ and at the best parameter set $\boldsymbol{\hat\theta}$.}
    \label{Fig_3}
\end{figure*}

\begin{figure*}[!ht]
    \centering
    \includegraphics[width=1.0\textwidth]{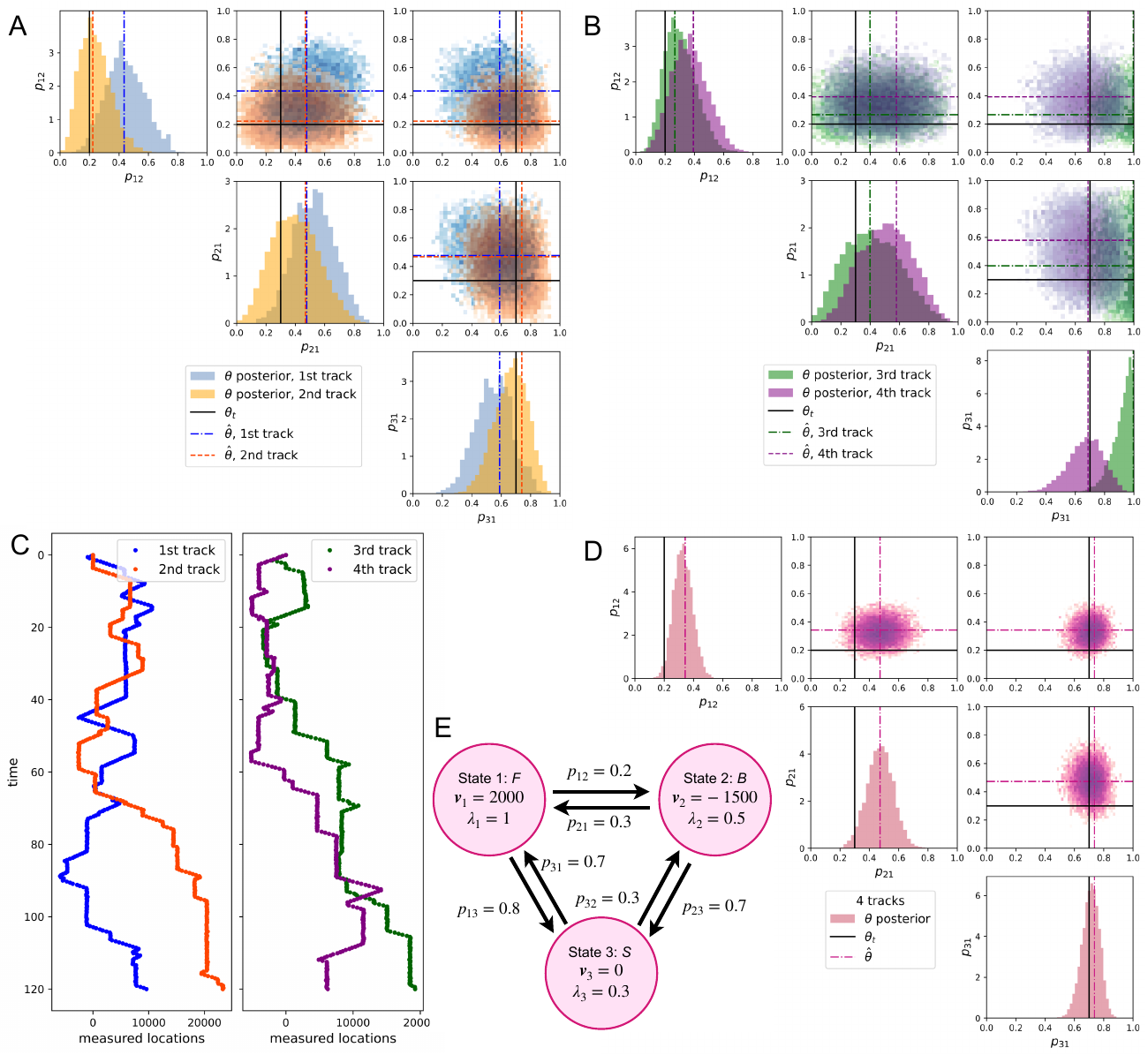}
    \caption{\textbf{Calibration of a three-state model to four in-silico data tracks separately and together}. \textbf{A} and \textbf{B} compare the univariate and bivariate distributions of the posterior of the parameters $p_{12},p_{21},p_{31}$ for the four data tracks in \textbf{C}. The parameter set estimated is $\boldsymbol{\theta}=[v_1,v_2,\log(\lambda_1), \log(\lambda_2),\log(\lambda_3),p_{12},p_{21},p_{31},\sigma]$ (posterior distributions for all parameters are shown in Supplementary Information Figures~S7 and S8). \textbf{C} shows the four data tracks used, collected with measurement noise $\sigma=50$ with fixed time between measurements $\Delta t= 0.3$, for a time interval of $T=120$, giving $N=400$ subsequent location increments. \textbf{D} shows the posterior distributions of the parameters $p_{12},p_{21},p_{31}$ using the four tracks together (posterior distributions for all parameters are shown in Supplementary Information Figure~S9). The black lines indicate the true parameters $\boldsymbol{\theta}_t$ specified in \textbf{E}, and the dashed lines indicate the best parameter sets $\boldsymbol{\hat\theta}$.}
    \label{Fig_4}
\end{figure*}

In this section, we use the framework to calibrate the parameters of three-state models, which, in addition to the forward and backward states in the two-state model (Figure~\ref{Fig_2}\textbf{B}), include a stationary state with fixed velocity $v_3=0$, and a switching rate $\lambda_3$ (Figure~\ref{Fig_3}\textbf{B} and Figure~\ref{Fig_4}\textbf{E}). Moreover, for models with at least three states we need to specify the network probability matrix with the property that the sum of the elements in each row must be unity. For three-state models the matrix has the form
\begin{equation}\label{Eq:P_matrix}
    \boldsymbol{P}:=
\begin{bmatrix}
    0 & p_{12} & 1-p_{12}
    \\
    p_{21} & 0 & 1-p_{21}
    \\
    p_{31} & 1-p_{31} & 0
\end{bmatrix}.
\end{equation}

Firstly, we study a simplified model in which we assume that there is no direct switching from the forward to the backward state, and vice versa. In this case $p_{12}=p_{21}=0$ and we only need to estimate $p_{31}$ (Figure~\ref{Fig_3}\textbf{B}). For a short track with $N=200$ data points (the blue dots in Figure~\ref{Fig_3}\textbf{C}), the framework can be used to calibrate the model to the data, however, some parameters cannot be confidently estimated. In particular, the posterior for $p_{31}$ is wide, with support across the entire prior interval $[0,1]$, indicating that the data is not sufficient to obtain a precise estimate of this probability (blue histograms in Figure~\ref{Fig_3}\textbf{A}). For a significantly longer track with $N=1600$ data points (Figure~\ref{Fig_3}\textbf{C}) the uncertainty for all the parameter estimates is reduced, including for the probability $p_{31}$ (orange histograms in Figure~\ref{Fig_3}\textbf{A}).

We further investigate the identifiability of the parameters of the network probability matrix, extending the results for a single probability parameter in Figure~\ref{Fig_3}, by considering a three-state model which includes the possibility of going directly from the forward state to the backward state and vice versa. In particular, we focus on studying the variability in the probability parameter estimates across datasets. We generate four different tracks with $N=400$ data points each (Figure~\ref{Fig_4}\textbf{C}) by setting the probabilities $p_{12}=0.2$, $p_{21}=0.3$ and $p_{31}=0.7$, which uniquely define the probability matrix, as in Equation~\eqref{Eq:P_matrix}, and all other model parameters as before (specified in Figure~\ref{Fig_4}\textbf{E}). We use each dataset to estimate all the model parameters, only fixing the stationary state velocity to zero. We calibrate the model to each track separately and to the four tracks together to investigate the uncertainty and variability in parameter posteriors.

Next, all three probability parameters are estimated. We observe that the chosen number of data points for each track ($N=400$) leads to wide estimated posteriors of the probability parameters  
(Figure~\ref{Fig_4}\textbf{A}-\textbf{B}) due to the low number of switches that occur in each data track. There is also a high variability between these posteriors across the four different tracks which reflects the variability in the data (Figure~\ref{Fig_4}\textbf{A}-\textbf{B}). From the third track alone, we cannot exclude the possibility that $p_{31}=1$, as its posterior is skewed towards one, which would correspond to a network in which the agent is only able to switch from the stationary state to the forward state. In contrast, all other parameters can be estimated with relatively low variability in the posterior compared to their prior for all four datasets (Supplementary Information Figures~S7 and S8).

Multiple tracks can be used together to reduce the uncertainty in parameter estimates and more accurately recover the true parameter set $\boldsymbol{\theta}_t$. In particular, the probability parameters posterior is reduced by a factor of around two when model calibration is performed using all four tracks together, rather than using the tracks individually (Figure~\ref{Fig_4}\textbf{D}). The posterior for the estimated parameter set is shown in Supplementary Information Figure~S9.

\subsection{Selection criteria choose a simplified network structure for a three-state model using data with no or rare direct switching between pairs of states}\label{Subsec:Results_network_selection}

\begin{figure*}[!ht]
    \centering
    \includegraphics[width=1.0\textwidth]{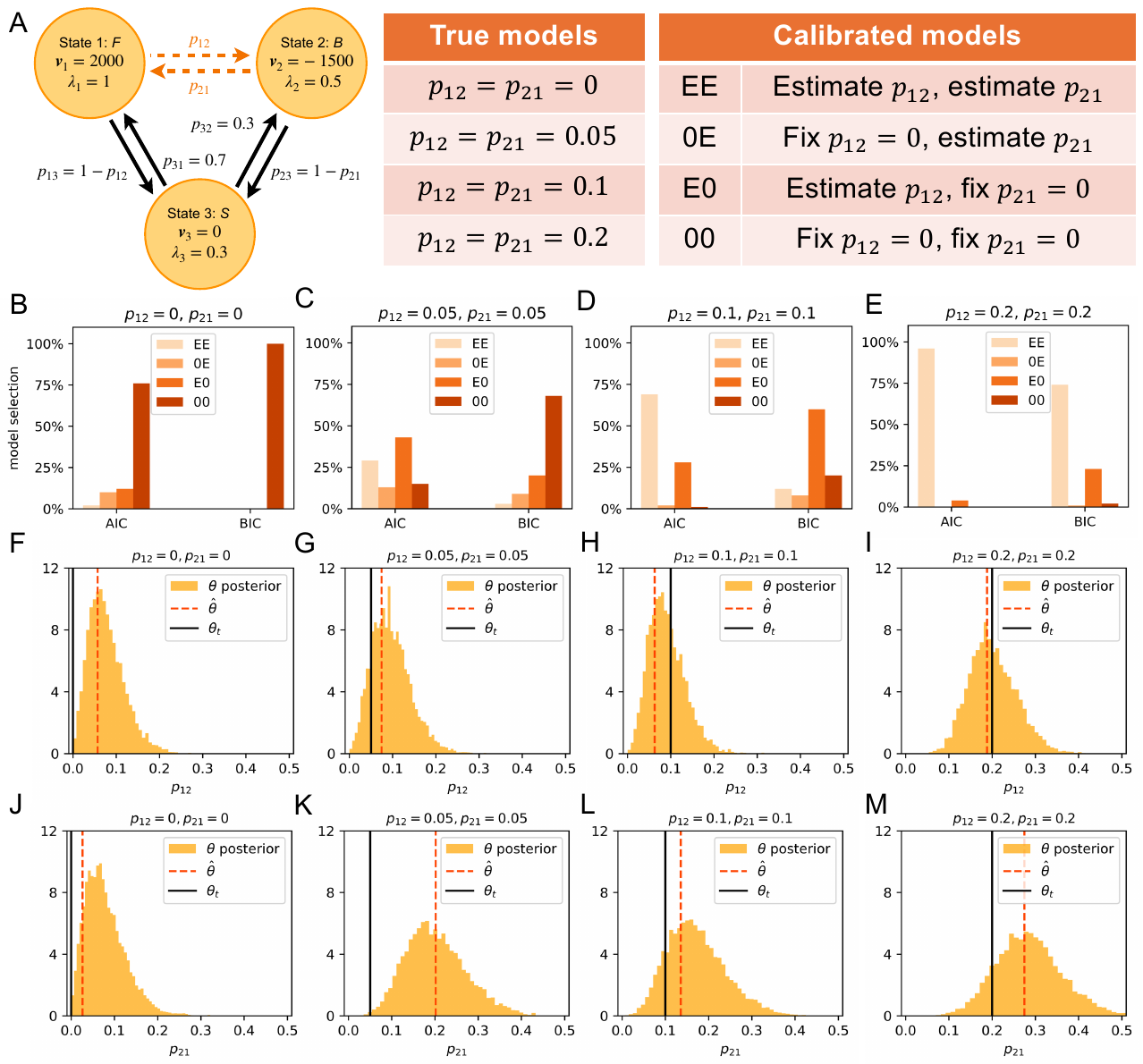}
    \caption{\textbf{Network selection for three-state models with zero and close-to-zero probabilities of switching from forward to backward motion and vice versa}. \textbf{A} specifies the three-state model parameters. In particular, the probabilities of switching from forward to backward motion and vice versa are set to $p_{12}=p_{21}=0$, $p_{12}=p_{21}=0.05$, $p_{12}=p_{21}=0.1$ and $p_{12}=p_{21}=0.2$. The measurement noise parameter is set to $\sigma=50$ with fixed time between measurements $\Delta t=0.3$, for a time interval $T=480$, corresponding to $N=1600$ subsequent location increments. For each model type we generated 100 datasets, and applied the framework using the track likelihood $P_1$ for each dataset separately. Four different models were calibrated: in model EE we estimate $p_{12}$ and $p_{21}$, in model 0E we fix $p_{12}=0$ and estimate $p_{21}$, in model E0 we estimate $p_{12}$ and fix $p_{21}=0$ and in model 00 we fix $p_{12}=0$ and $p_{21}=0$. The full parameter set estimated is $\boldsymbol{\theta}=[v_1,v_2,\log(\lambda_1), \log(\lambda_2),\log(\lambda_3),p_{12},p_{21},p_{31},\sigma]$ (for model EE). \textbf{B}-\textbf{E} compare the model selected according to the lowest AIC and BIC. \textbf{F}-\textbf{M} posteriors of $p_{12}$ and $p_{21}$, respectively, obtained calibrating model EE using a single data track generated for all models. The MCMC chain were initialised at the true parameter set used to generate the data, to allow for quicker convergence, with the exception of the probability parameters which are sampled uniformly in $[0,1]$.}
    \label{Fig_5}
\end{figure*}

In Figure~\ref{Fig_4} we estimated the posterior of the probability parameter $p_{31}$ from datasets generated using a true $p_{31}=0.7$. For the third track (Figure~\ref{Fig_4}\textbf{B}) we estimated $p_{32}$ to be close to zero, which may indicate the lack of direct switching from state $s=3$ to state $s=2$. This leads us to consider the question of model selection to determine the network connecting the model states. As such, in this section we characterise the nature of the network of states of three-state models (Figure~\ref{Fig_5}\textbf{A}) focusing on direct switching between the forward state $s=1$ and the backward state $s=2$. Using the AIC (Equation~\eqref{Eq:AIC}) and BIC (Equation~\eqref{Eq:BIC}), we test when direct switching should be included in the model, calibrated using tracks of $N=1600$ data points. We consider 100 datasets generated from four three-state models, with increasing probabilities $p_{12}=p_{21}$ in the set $\{0, 0.05, 0.1, 0.2\}$ and we use them to calibrate four three-state models with distinct network structures specified in Figure~\ref{Fig_5}\textbf{A} (calibrated models EE, 0E, E0 and 00). In all calibrated models we estimate the non-zero velocities, the switching rates, the probability $p_{31}$ and the noise parameter. In model EE we additionally estimate both $p_{12}$ and $p_{21}$; in model 0E we fix $p_{12}=0$ and estimate $p_{21}$; in model E0 we estimate $p_{12}$ and fix $p_{21}=0$; and in model 00 we fix $p_{12}=p_{21}=0$.

Firstly, we consider model calibration with data generated with probabilities $p_{12}=p_{21}=0$. The AIC usually selects model 00 (76\%), which fixes $p_{12}=p_{21}=0$, and the BIC always selects model 00 (100\%) (Figure~\ref{Fig_5}\textbf{B}). In this case, both criteria hint towards selecting model 00, suggesting that the improvement in the likelihood estimate in other models is not sufficient to motivate the calibration of additional parameters. As expected, even when these probabilities are estimated, the majority of the masses of their posteriors are concentrated towards zero (e.g. the estimated distributions for $p_{12}$ and $p_{21}$ in model EE in Figure~\ref{Fig_5}\textbf{F} and \textbf{J} for the first data track).

Secondly, we consider model calibration using data generated with $p_{12}=p_{21}=0.05$. The AIC selects model E0 more frequently than other models (43\%), while the BIC is more consistent, often selecting model 00 (68\%) (Figure~\ref{Fig_5}\textbf{C}). Here, we see that, as the true probability parameters are now non-zero, the criteria become less likely to choose the simplest model 00. To test this further we consider model calibration with data generated with higher probabilities $p_{12}=p_{21}=0.1$. In this case, the AIC often selects model EE (69\%) in which the two probabilities are estimated, while the BIC often selects model E0 (60\%), estimating only $p_{12}$ (Figure~\ref{Fig_5}\textbf{D}). Given that the true $p_{12}$ and $p_{21}$ are both $0.1$, the choice of the BIC of fixing $p_{21}=0$ but estimating $p_{12}$ is a consequence of the other model parameters. Indeed, the backward state is visited less often than the forward state after the stationary state as the true $p_{31}=0.7$. It follows that the number of switches from the backward state (to the forward or stationary states) is lower than the number of switches from the forward state (to the backward or stationary states). Hence, this leads to a lower variation in the maximum likelihood estimation from the choice of $p_{21}$ than of $p_{12}$. This disproportion in the number of switches is also reflected in a higher variability when estimating $p_{21}$ compared to estimating $p_{12}$ (Figure~\ref{Fig_5}\textbf{G} versus \textbf{K} for $p_{12}=p_{21}=0.05$ and \textbf{H} versus \textbf{L} for $p_{12}=p_{21}=0.1$ for a single track).

Finally, we consider model calibration with data generated with probabilities $p_{12}=p_{21}=0.2$. In this case, both the AIC and the BIC usually select model EE (96\% and 74\%; Figure~\ref{Fig_5}\textbf{D}), and in this case the AIC is more consistent than the BIC. Hence, as the true probabilities get further away from zero, the impact of incorporating direct switching in the calibrated model on the likelihood estimate increases, despite the uncertainty in estimating the related probability parameters.

We also note that the AIC and BIC model choices are likely to agree when the true parameters are $p_{12}=p_{21}=0$ or $p_{12}=p_{21}=0.2$ (76\% and 78\%, respectively), while when $p_{12}=p_{21}=0.05$ and $p_{12}=p_{21}=0.1$ the model choices tend to disagree on the selected model (agree in 30\% and 32\% of the cases, respectively). Overall, the results suggest that a sensible strategy when selecting between models for a single data track is to always compute both criteria. If the criteria agree, they provide supporting evidence for the chosen model network. If they disagree, additional data should be collected to further evaluate the network choice, either by obtaining a longer track or by combining multiple tracks.

\subsection{Selection criteria typically choose models in which each state is associated with a distinct velocity}\label{Subsec:Results_model_selection}

\begin{figure*}[!ht]
    \centering
    \includegraphics[width=1.0\textwidth]{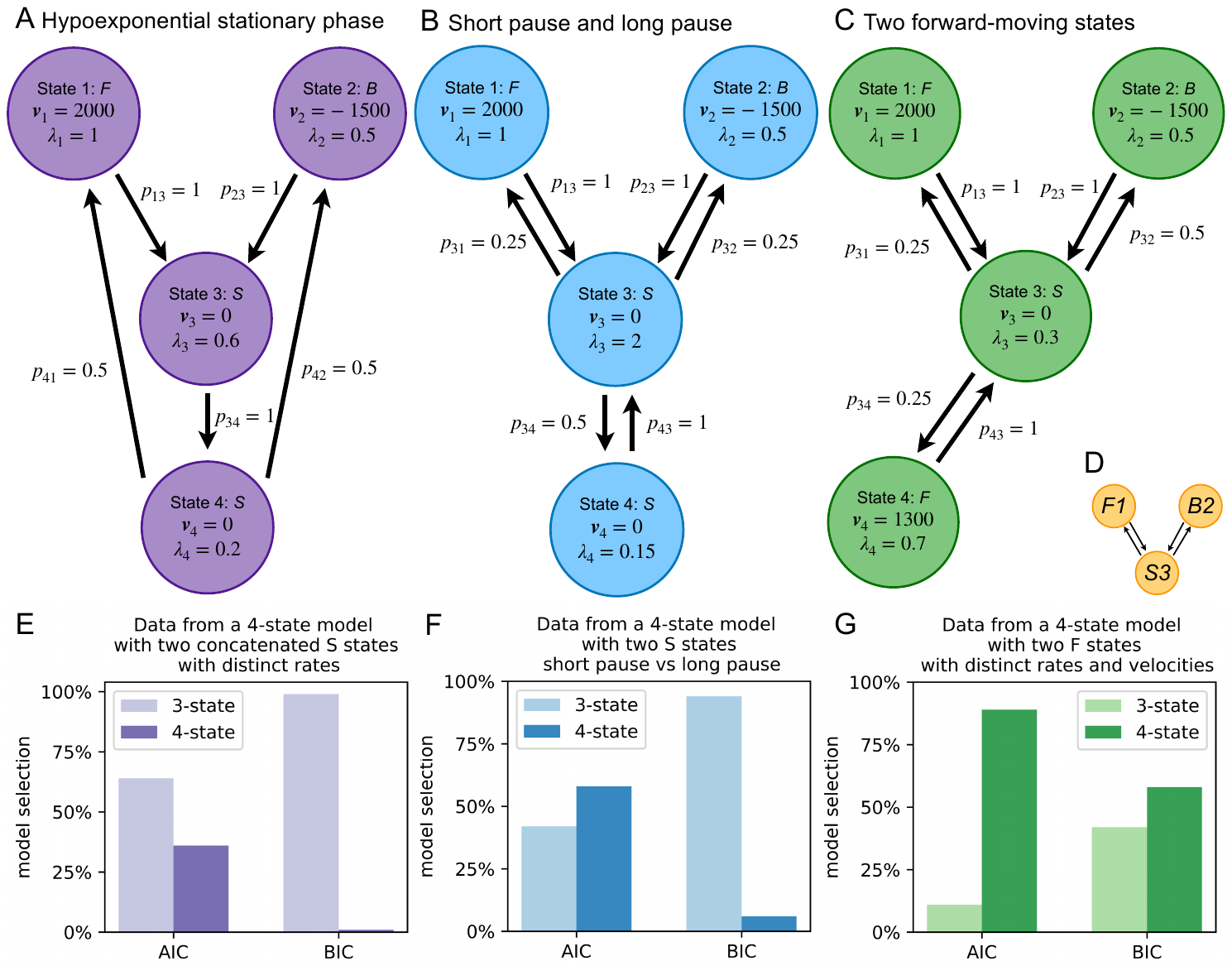}
    \caption{\textbf{Selecting between a three-state model and a four-state model for data generated using three distinct four-state models}. \textbf{A}-\textbf{C} specify the parameters and structure of three four-state models used to generate 100 in-silico data tracks. The data tracks are collected with measurement noise $\sigma=50$ with fixed time between measurements $\Delta t=0.3$, for a time interval of $T=120$, which gives $N=400$ subsequent location increments for each dataset. \textbf{D} specifies the three-state model calibrated using the datasets generated from each four-state model. The parameter set calibrated for model \textbf{A} is $\boldsymbol{\theta}=[v_1,v_2,\log(\lambda_1), \log(\lambda_2),\log(\lambda_3),\log(\lambda_4),p_{41},\sigma]$, for model \textbf{B} is $\boldsymbol{\theta}=[v_1,v_2,\log(\lambda_1), \log(\lambda_2),\log(\lambda_3),\log(\lambda_4),p_{31},p_{32},\sigma]$, for model \textbf{C} is $\boldsymbol{\theta}=[v_1,v_2,v_4,\log(\lambda_1), \log(\lambda_2),\log(\lambda_3),\log(\lambda_4),p_{31},p_{34},\sigma]$, and for model \textbf{D} is $\boldsymbol{\theta}=[v_1,v_2,\log(\lambda_1), \log(\lambda_2),\log(\lambda_3),p_{31},\sigma]$. The MCMC chain were initialised at the true parameter set used to generate the data, to allow for quicker convergence, and the burnin is reduced to 1000 iterations. \textbf{E}-\textbf{G} show the choice between the three-state model and the four-state model for each of the 100 datasets tested according to the AIC and BIC.}
    \label{Fig_6}
\end{figure*}

Here, we aim to expand on the question of model selection, already considered in Section~\ref{Subsec:Results_network_selection}, by comparing models with a varying number of states. Specifically, we generate 100 in-silico datasets of $N=400$ data points using three distinct four-state models (Figure~\ref{Fig_6}\textbf{A}-\textbf{C}) which share a forward, a backward and a stationary state, but vary in the fourth state. The state $s=4$ is stationary in models \textbf{A} and \textbf{B}, which differ in network structures, while in model \textbf{C} it is a forward state with a distinct velocity from state $s=1$. We use the datasets generated using these models to calibrate the four-state models themselves and to calibrate the three-state model in Figure~\ref{Fig_6}\textbf{D}. The results obtained allow us to assess the importance of including the fourth state in different scenarios and to discuss the identifiability of the model parameters.

\begin{figure*}[!ht]
    \centering
    \includegraphics[width=1\textwidth]{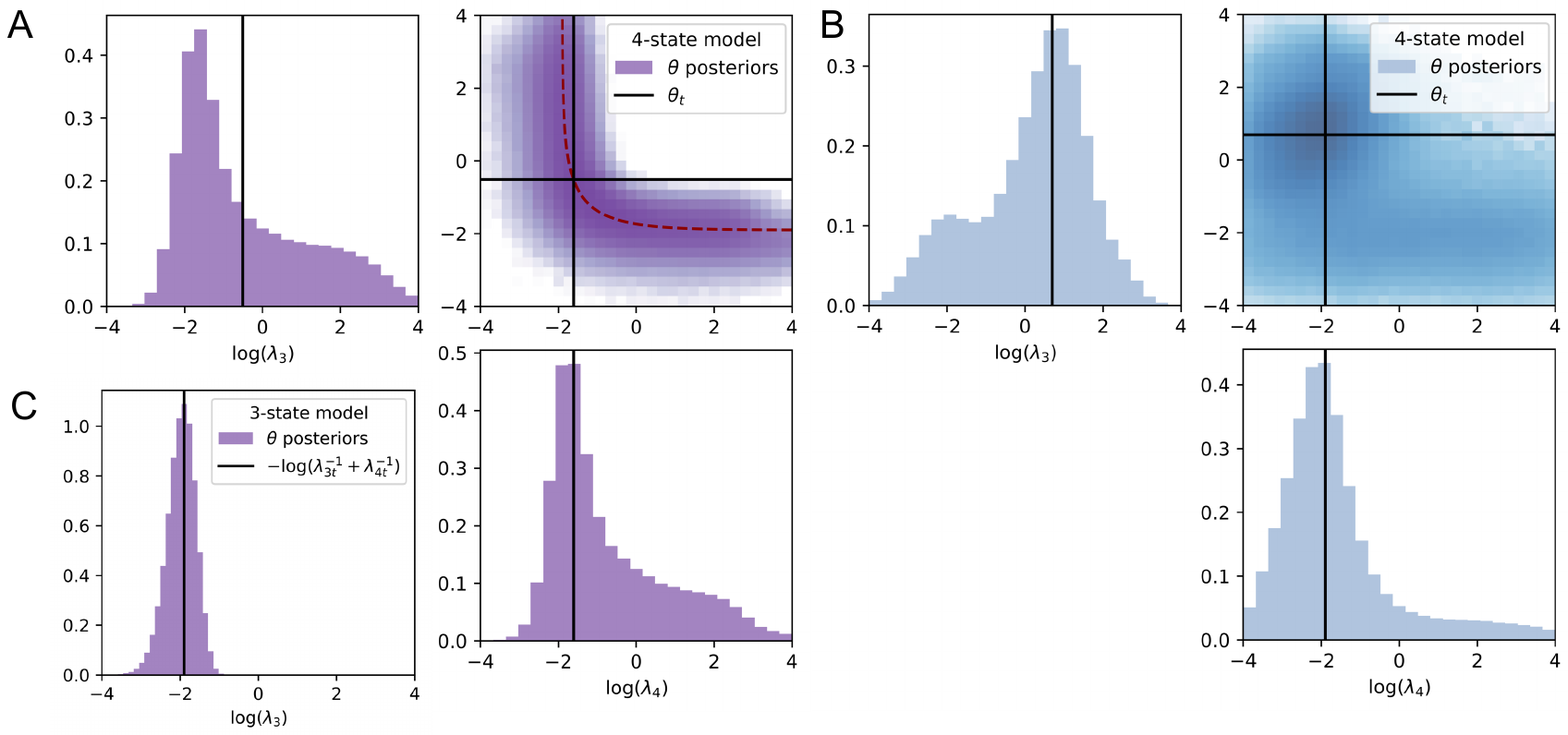}
    \caption{\textbf{Comparison between the mixture of posteriors for the calibrated four-state model versus three-state model in Figure~\ref{Fig_6} with horizontal axes corresponding to the priors used}. \textbf{A} shows the univariate and bivariate posteriors for the parameters $\log(\lambda_3)$ and $\log(\lambda_4)$ obtained calibrating the four-state model in Figure~\ref{Fig_6}\textbf{A}, and the dashed red line represents the equality $\lambda_3^{-1} +\lambda_4^{-1}=\lambda_{3t}^{-1} +\lambda_{4t}^{-1}$. \textbf{B} shows the univariate and bivariate posteriors for the parameters $\log(\lambda_3)$ and $\log(\lambda_4)$ obtained calibrating the four-state model in Figure~\ref{Fig_6}\textbf{B}. \textbf{C} shows the univariate posterior for the parameter $\log(\lambda_3)$ obtained calibrating the three-state model in Figure~\ref{Fig_6}\textbf{D} using the data generated by the four-state model in Figure~\ref{Fig_6}\textbf{A}.}
    \label{Fig_7}
\end{figure*}

Firstly, we consider the model in Figure~\ref{Fig_6}\textbf{A}, characterised by a second stationary state, $s=4$, visited after the stationary state $s=3$. Imposing this network is equivalent to imposing a hypoexponential distribution for the total time spent in a stationary state, instead of the exponential distribution which arises from considering a single stationary state. In this case, the three-state model is often selected by the AIC (64\%) and almost always selected by the BIC (99\%) (Figure~\ref{Fig_6}\textbf{E}) and it has one fewer parameter ($\lambda_4$) than the four-state model. Moreover, the parameters $\lambda_3$ and $\lambda_4$ are almost practically non-identifiable as their mixture posterior distributions, combined across the 100 datasets considered, are comparable (Figure~\ref{Fig_7}\textbf{A}). However, the average time spent in the stationary phase, obtained as the sum $\lambda_3^{-1} + \lambda_4^{-1}$, is identifiable (Figure~\ref{Fig_7}\textbf{A} dashed red line), and its inverse, $(\lambda_3^{-1} +\lambda_4^{-1})^{-1}$, corresponds to the estimated rate $\lambda_3$ in the three-state model with only one stationary state (Figure~\ref{Fig_7}\textbf{C}). We conclude that for the four-state model the hypoexponential stationary phase with two rates $\lambda_3$ and $\lambda_4$ is well approximated by the exponential distribution with rate $(\lambda_3^{-1} +\lambda_4^{-1})^{-1}$, and therefore the two stationary states can be substituted by a single stationary state as they do not significantly improve the model likelihood.

Next, we consider the model in Figure~\ref{Fig_6}\textbf{B} in which the fourth state is again a stationary state. The stationary state $s=3$ corresponds to a short pause in the motion due to the high switching rate $\lambda_3=2$, while the stationary state $s=4$ corresponds to a much longer pause in the motion ($\lambda_4=0.15$) and is only directly connected to state $s=3$. In this case, the AIC selects the four-state model (58\%) slightly more often than the three-state model; in contrast, the BIC reliably selects the three-state model (94\%) (Figure~\ref{Fig_6}\textbf{F}). Overall, the BIC is again more consistent than the AIC and prefers the three-state model which has two fewer parameters ($\lambda_4$ and $p_{32}$, now needed as $p_{34}$ is also present) compared to the four-state model. Moreover, for this model structure (Figure~\ref{Fig_6}\textbf{B}) it is possible to identify the switching rates $\lambda_3$ and $\lambda_4$ (Figure~\ref{Fig_7}\textbf{B}).

Finally, we consider the model in Figure~\ref{Fig_6}\textbf{C}, in which the fourth state is a forward state. The forward state $s=1$ has velocity $v_1=2000$ and switching rate $\lambda_1=1$, while the forward state $s=4$ has velocity $v_4=1300$ and switching rate $\lambda_4=0.7$. The AIC usually prefers the four-state model (89\%) and the BIC also selects the four-state model but less frequently (58\%) (Figure~\ref{Fig_6}\textbf{G}). Hence, having two states with distinct velocities motivates the use of one state for each velocity, despite the need to calibrate three additional parameters compared to the three-state model ($v_4$, $\lambda_4$ and $p_{34}$).

Overall, the selection criteria suggest that the additional parameters required with a higher number of states are often not justified by the improvement in the maximum likelihood estimate. The BIC is consistent in selecting the simpler model when multiple states with the same velocity are used, as they can often be simplified to a single state. In this case, we recommend using the simpler model, especially since some model structures may lead to significant uncertainty in parameter estimates (Figure~\ref{Fig_7}\textbf{A}). In contrast, when the additional state has a distinct velocity from the other states, the AIC quite consistently prefers the most complicated model (Figure~\ref{Fig_6}\textbf{C}), and this is also often selected by the BIC and further supported by the relative difference in the estimated noise parameter. As in Section~\ref{Subsec:Results_network_selection}, we recommend that, when selecting between models to describe a single data track, both criteria should be calculated, and, if the criteria disagree, more data or further considerations may be necessary to choose a model. For example, we suggest computing the relative difference in the estimated noise standard deviation obtained from a simpler model to a more complex model, calibrated to the same dataset. We interpret a higher noise parameter in the simpler model as indicative of model misspecification, as the data points that are poorly explained by the simpler model are instead described through an inflated noise term. We observe that the relative difference in the noise standard deviation is significant (up to $19\%$) for model \textbf{C}, with the highest values obtained using datasets for which the AIC and BIC select the four-state model (Supplementary Information Figure~S10). Hence, a significant relative difference in the noise parameter can further support the preference of the selection criteria of a more complex model.

\subsection{Application to the intracellular motion of mRNA in \textit{Drosophila} axons}\label{Subsec:Results_experimental_data}

\begin{figure*}[!ht]
    \centering
    \includegraphics[width=1\textwidth]{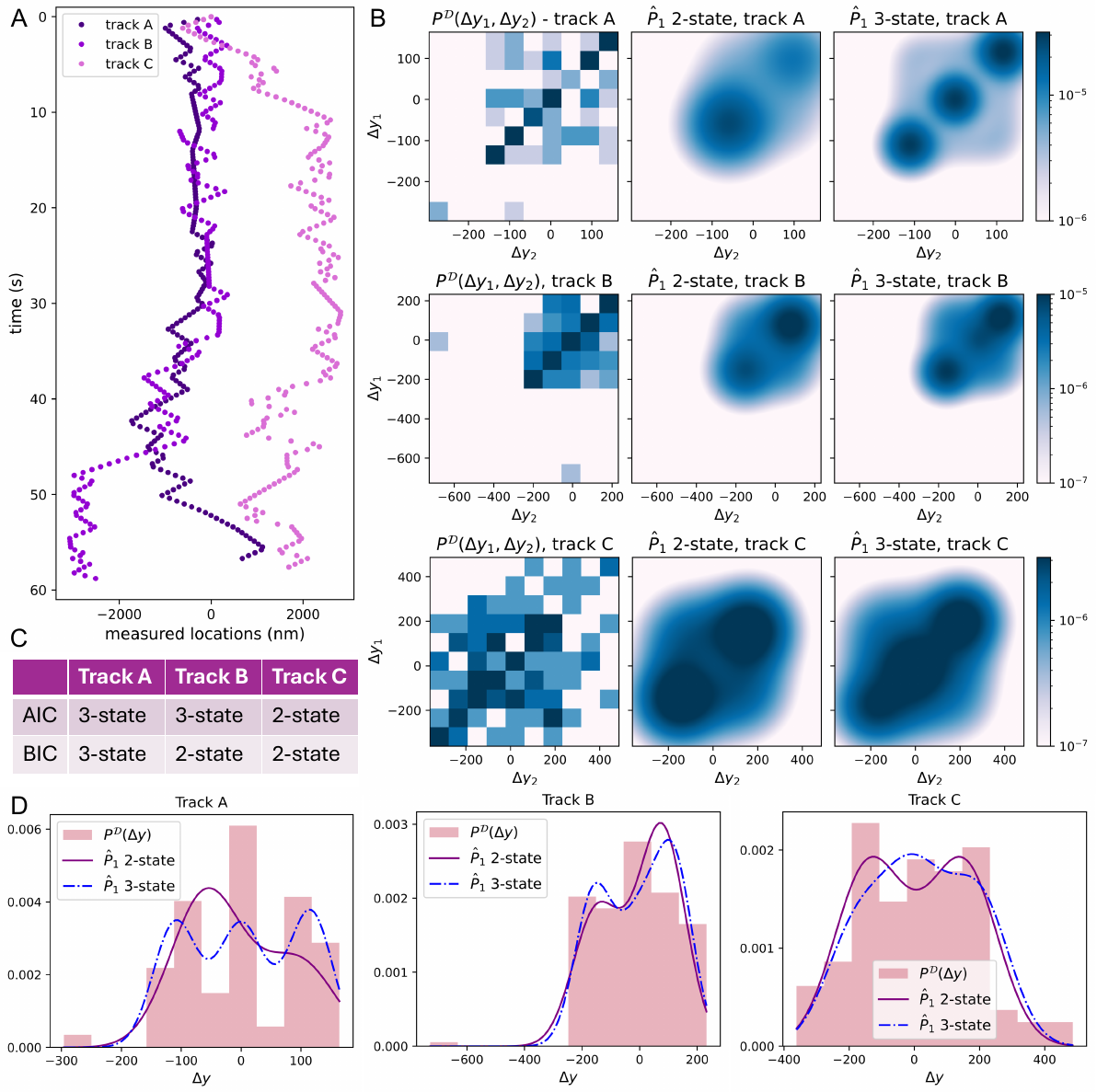}
    \caption{\textbf{Calibration of and selection between two-state and three-state models using mRNA tracks along \textit{Drosophila} neurons}. \textbf{A} shows tracks obtained from live imaging of \textit{Rab11} and \textit{Gdi} mRNAs by the MS2/MCP system. Track A (\textit{Gdi}) and tracks B and C (\textit{Rab11}) are obtained from cultured primary neurons from the brain of day 6-7 larvae, at day \textit{in vitro} 3, with fixed $\Delta t=0.3$s (for further details on the data collection see Chapter~4 in \citep{chaiamarit2023axonal}). Trajectories are extracted from live images using KymoAnalyzer \citep{neumann2017kymoanalyzer}. \textbf{B} compares the data distribution $P^{\mathcal{D}}$ for a set of two subsequent location increments $[\Delta y_1, \Delta y_2]$ to the likelihoods at the best parameter set $\hat P_1$ obtained from the calibration of the two-state model and the three-state model for each track. \textbf{C} shows the model selected according to the lowest AIC and BIC for each track. \textbf{D} compares the data distribution $P^{\mathcal{D}}$ for a single location increment $\Delta y$ to the likelihoods at the best parameter set $\hat P_1$ obtained from the calibration of the two-state model and the three-state model for each track.}
    \label{Fig_8}
\end{figure*}

In this section, we apply the framework to interpret experimental data that track the motion of mRNA along the axons of \textit{Drosophila} neurons. In Figure~\ref{Fig_8}\textbf{A} we show experimental data tracks of \textit{Gdi} (Rab GDP dissociation inhibitor) (track A) and of \textit{Rab11} (tracks B and C) moving along \textit{Drosophila} neurons. These tracks show the alternation of motion in forward and backward directions, and, particularly in tracks A and B, we notice some vertically-aligned dots that may suggest the presence of stationary phases (Figure~\ref{Fig_8}\textbf{A}). We highlight that the stochastic velocity-jump models in one spatial dimension are suitable to describe the typical motion of mRNA along the axons of the neurons, characterised by the alternation of directed anterograde (forward state) and retrograde (backward state) motion and by the potential presence of stationary phases (stationary state). We answer biological questions, such as whether pausing phases are present, by calibrating and selecting between
two-state models, with a forward and a backward state, and three-state models with an additional stationary state.

Firstly, by calibrating the three-state model we see that, for all three tracks, the switching between movements in opposite directions can be instantaneous. Indeed, the probability parameters $p_{12}$ (from forward to backward motion) and $p_{21}$ (from backward to forward motion) are estimated to be close to one, indicating that direct switching between the forward state ($s=1$) and the backward state ($s=2$) is prevalent and that the stationary state ($s=3$) is infrequently visited (Supplementary Information Figures~S12, S14 and S16). Biologically, the direct switching between forward and backward states may indicate that while the mRNA is located close to the microtubules and bound to the molecular motors it alternates anterograde and retrograde motion. Moreover, the presence of a stationary state indicates phases of no significant movement, which can occur when the mRNA is not bound to motors.

We test the presence of a stationary phase by performing model selection. The AIC and BIC both select the three-state model for track A (\textit{Gdi}), and the two-state model for track C (\textit{Rab11}), while they disagree for track B (\textit{Rab11}) (Figure~\ref{Fig_8}\textbf{C}). The preference of the three-state model for the \textit{Gdi} track A could be anticipated as there are three high peaks in the data distribution (at around $-100\text{nm}$, $0\text{nm}$ and $100\text{nm}$ in Figure~\ref{Fig_8}\textbf{D}), while there are less distinguishable peaks for the \textit{Rab11} tracks B and C (Figure~\ref{Fig_8}\textbf{D}). The selection of the three-state model for \textit{Gdi} suggests that its motion comprises of stationary phases, while for \textit{Rab11} there is no strong evidence of the presence of a stationary state as the BIC selects the two-state model for track C, while the criteria disagree for track B.

Finally, for all tracks, the estimated noise standard deviation is lower for the three-state model compared to the two-state model (Supplementary Information Figures~S11-S16). In particular, we note that the noise parameter $\sigma$ estimated calibrating the two-state model is reduced by $37\%$, $19\%$ and $8\%$ for tracks A, B and C, respectively, calibrating the three-state model (Supplementary Information Figures~S11-S16). Similarly to Section~\ref{Subsec:Results_model_selection}, the higher noise standard deviation estimate in the simpler model may indicate model misspecification. Hence, the relative difference in noise standard deviation is most significant (track A, $37\%$) when the AIC and BIC both prefer the three-state model, further justifying their model selection and, therefore,  presence of the stationary state for the \textit{Gdi} track.

\section{Discussion and conclusions}\label{Sec:Discussion_and_conclusions}

In this manuscript, we present a likelihood-based Bayesian inference framework to calibrate a general one-dimensional $n$-state stochastic velocity-jump model to single-agent tracking data. The calibration framework provided is computationally efficient, due to the use of likelihood approximations corresponding to the model solutions subject to discrete-time noisy observations. We overcome the challenge of estimating the parameters of a continuous model using data collected at discrete time steps and with measurement noise.

Firstly, we considered a simple two-state model with a forward and a backward state and investigated whether the framework can be used to recover the model parameters from in-silico data (Figure~\ref{Fig_2}). We conclude that in this case all model parameters can be recovered given that the data are sampled sufficiently frequently compared to the state-switching rates. An indication that the data are collected sufficiently frequently to capture the underlying process is the presence of peaks in the data distribution, corresponding to the different model states (e.g. Figure~\ref{Fig_2}\textbf{D}).

Next, we tested the ability of the framework to recover the parameters of a three-state model with a forward, a backward, and a stationary state, including inferring the network probability matrix, given a sufficient data collection frequency. Our results demonstrate that, while the velocities, rates, and noise parameters can be estimated with low variability in their posterior compared to their prior, the probabilities in the network matrix often show a posterior variability comparable to their prior (Figure~\ref{Fig_3}). The uncertainty in all parameter estimates can be reduced by increasing the number of data points in a track (Figure~\ref{Fig_3}) or by using a number of tracks from the same model together to estimate the likelihood (Figure~\ref{Fig_4}).

We also tackled the question of model selection using the AIC and the BIC. Firstly, we focused on selecting between three-state models with different networks connecting the states. We generated 100 data tracks for four models with increasing network probabilities $p_{12}$ and $p_{21}$, and we used them to calibrate four model types with distinct networks (Figure~\ref{Fig_5}\textbf{A}). The BIC consistently selects the simplest model (with the probabilities $p_{12}$ and $p_{21}$ set to zero) when these probabilities in the true model are zero or very close to zero (Figure~\ref{Fig_5}\textbf{B}-\textbf{C}). In contrast, as the probabilities get further from zero, the AIC more consistently selects the more complex fully-connected model (Figure~\ref{Fig_5}\textbf{E}). Hence, when selecting between models, we recommend the computation of both criteria. For a single track, if the two criteria agree on the model selection, then the choice is likely consistent across datasets. However, if they disagree, their choice is likely not consistent and a comparison across dataset could be carried using multiple tracks when available.

We then focused on the question of selecting between models with a different number of states using the AIC and BIC. We generated 100 data tracks from three different four-state models and used them to calibrate the respective four-state model and a simplified three-state model (Figure~\ref{Fig_6}\textbf{A}-\textbf{D}). In this case, the BIC very consistently prefers the simplest model for data generated with two stationary states (Figure~\ref{Fig_6}\textbf{E}-\textbf{F}), while the AIC more consistently prefers the four-state model when two forward states with different velocities are present (Figure~\ref{Fig_6}\textbf{G}), also supported by the significant relative difference in estimated noise standard deviation.

Moreover, we highlight that there may be significant uncertainty in some parameter estimates for four-state models with a specific type of structure. In particular, in a model with two concatenated states with the same velocity (Figure~\ref{Fig_6}\textbf{A}), the switching rates associated with those states may not be practically identifiable (Figure~\ref{Fig_7}\textbf{A}). However, we also note that for a model with a different network structure (Figure~\ref{Fig_6}\textbf{B}) these rate parameters can be identified (Figure~\ref{Fig_7}\textbf{B}).

We also applied our framework to calibrate two-state and three-state models to experimental data tracking mRNA motion in \textit{Drosophila} neurons. We found that the three-state model with a forward, a backward and a stationary state better captures the behaviour of \textit{Gdi} as supported by the AIC and BIC and the significant relative difference in estimated noise standard deviation compared to the two-state model with a forward and a backward state. In contrast, to describe the motion of \textit{Rab11} the two-state model is usually preferred. From the model calibration and selection we also obtained further insights on the mRNA motion in \textit{Drosophila} neurons, such as that the switching between forward and backward motion can happen instantaneously.

The framework provided allows for efficient parameter inference of stochastic velocity-jump models, offering several advantages with respect to the particle filtering pseudomarginal methods generally used to estimate parameters of partially-observed Markov processes~\citep{simpson2022reliable, warne2020practical, king2016statistical, andrieu2010particle}. In particle filtering methods, a number of particles is specified and their trajectories are simulated and weighted to obtain a log-likelihood approximation. A challenge of these methods is that they require choosing a number of particles which guarantees that at least one trajectory remains sufficiently close to the data to avoid obtaining a degenerate likelihood (Supplementary Information Figure~S17\textbf{A}). Moreover, the log-likelihood approximation in our framework is deterministic, while it is variable using particle filtering methods due to their stochasticity (Supplementary Information Figure~S17\textbf{C}), which can make the comparison between parameter sets less consistent and result in slower convergence. An advantage of our framework is that it does not face these issues. Finally, we highlight that, for the two-state model in Figure~\ref{Fig_2}\textbf{B}, computing the log-likelihood with the track likelihood approximation is around five times faster than with a particle filtering method for $N_p=100$ particles (Supplementary Information Figure~S17\textbf{B}). Moreover, we note that the runtime for the particle filtering method increases proportionally with the number of particles $N_p$.

We use a simple implementation of the Metropolis-Hastings MCMC algorithm. However, the likelihood approximations are suitable to be used within any other likelihood-based inference method, potentially allowing for improved speed of convergence. When the exact posterior is required, the proposed framework could be used to compute an approximate posterior, then, particle filtering methods could be applied to obtain the exact posterior using the approximate posterior obtained to inform the prior to enable faster convergence.

This work could also inspire extensions to frameworks for the calibration of other partially-observed Markov models using discrete and noisy data. Some biological systems are described by the alternation of directed movement and diffusion. The motion of cargoes within a neuron is known to be directed, characterised by jumps in velocity, when cargoes are being transported as they are strongly bound to molecular motors, while it is primarily diffusive when they are weakly bound to motors or are searching for binding sites~\citep{gennerich2009walking}. Diffusive states may be incorporated in the models presented and in the likelihood approximations; the framework could be updated to be used to calibrate and select between these models to study the roles of diffusive and directed motion from noisy data.

In conclusion, we provide a framework to calibrate $n$-state velocity-jump models designed to describe single-agent motion in one spatial dimension, based on approximations of the model solutions subject to discrete-time noisy data. Moreover, our work paves the way for the construction of similar inference frameworks for other stochastic models.


\section*{Declarations}

\bmhead*{Funding}
A.C. is supported by an EPSRC/UKRI Doctoral Training Award. A.P.B. thanks the Mathematical Institute for a Hooke Research Fellowship. T.C. was supported by a Scripps-Oxford PhD Scholarship. I.D. was supported by a Wellcome Investigator Award (209412/Z/17/Z) and is supported by funding from the University of Glasgow, associated with his Strategic Professorship. R.E.B. is supported by a grant from the Simons Foundation (MP-SIP-00001828). For the purpose of open access, the author has applied a CC BY public copyright licence to any author accepted manuscript arising from this submission.

\bmhead*{Conflict of interest}
The authors declare that they have no conflict of interest.


\bmhead*{Consent for publication}
All the authors approved the final version of the manuscript.

\bmhead*{Code availability}
The code is available on GitHub in the repository\\ \href{https://github.com/a-ceccarelli/inference_n-state_VJ_model}{a-ceccarelli/inference\_n-state\_VJ\_model}.

\bmhead*{Data availability}
The data is available on GitHub in the repository\\ \href{https://github.com/a-ceccarelli/inference_n-state_VJ_model}{a-ceccarelli/inference\_n-state\_VJ\_model}.


\bmhead*{CRediT author statement}
A.C.: Writing - Original Draft, Writing - Review and Editing, Conceptualisation, Methodology, Formal Analysis, Visualisation, Software Implementation. T.C.: Investigation (Experimental Data Collection). I.D.: Resources (Experimental Data Collection). A.P.B and R.E.B.: Supervision, Conceptualisation, Writing - Review and Editing.


\bibliography{sn-bibliography}

\end{document}


\title{\centering \textbf{Supplementary Information} \\
A likelihood-based Bayesian inference framework \\ for the calibration of and selection between \\ stochastic velocity-jump models}

\author*[1]{\fnm{Arianna} \sur{Ceccarelli}\orcidlink{0000-0002-9598-8845}}\email{ceccarelli@maths.ox.ac.uk}

\author[2]{\fnm{Alexander P.} \sur{Browning}\orcidlink{0000-0002-8753-1538}}

\author[3,4]{\fnm{Tai}\sur{Chaiamarit}}

\author[3,5]{\fnm{Ilan}\sur{Davis}\orcidlink{0000-0002-5385-3053}}

\author[1]{\fnm{Ruth E.} \sur{Baker}\orcidlink{0000-0002-6304-9333}}

\affil[1]{\centering
Mathematical Institute, University of Oxford, UK}
\affil[2]{\centering
School of Mathematics and Statistics, University of Melbourne, Australia}
\affil[3]{Department of Biochemistry, University of Oxford, UK}
\affil[4]{Department of Physiology, Faculty of Science, Mahidol University, Thailand}
\affil[5]{School of Molecular Biosciences, University of Glasgow, UK}

\maketitle

\renewcommand\thefigure{S\arabic{figure}}
\renewcommand\thesection{S\arabic{figure}}
\setcounter{figure}{0}

\section*{S1 Details on the Bayesian inference framework}\label{SI:S1}

In order to allow faster convergence of the posterior, we use the burnin phase to operate a first parameter search while adaptively updating the covariance matrix. We start with the same covariance matrix for all chains, specified in Supplementary Information Section S2. We set the number of adaptive updates $n_A$ for each chain, a number of samples to discard $b$ (burnin), an initial covariance matrix $\boldsymbol{\Sigma}_0$, the data $\mathcal{D}$, a proposal density $q=q(\boldsymbol{\theta}, \boldsymbol{\Sigma})$.

\begin{algorithm}
\caption{Bayesian inference framework}
\begin{algorithmic}[1]
\State Given $q,\boldsymbol{\Sigma}_0, \boldsymbol{\theta}_0, n_A, b, \hat n$
\State Set split burnin $s=b/n_A$
\State Set $k$ as the size of $\boldsymbol{\theta}_0$
\State Set $\boldsymbol{\Sigma}=\boldsymbol{\Sigma}_0$
\For{$i =1,2,\ldots,n_A$}
    \State Call the Metropolis-Hastings MCMC \State $\boldsymbol{\Theta}, \boldsymbol{\log \pi}=\text{MetropolisHastings}(q=q,\ \boldsymbol{\Sigma}=\boldsymbol{\Sigma},\ \boldsymbol{\theta}_0=\boldsymbol{\theta}_0,\ \hat n = s_b$)
    \State Update the covariance matrix $$\boldsymbol{\Sigma}=\frac{2.38^2}{k}\text{COV}(\boldsymbol{\Theta}), \text{ where }\text{COV}(\cdot) \text{ denotes the covariance matrix}$$
    \State Start the chain at the last iteration $\boldsymbol{\theta}_0=\boldsymbol{\Theta}_{s,:}$
\EndFor
\State Run the Metropolis-Hastings MCMC for the remaining iterations
\State $\boldsymbol{\Theta}, \boldsymbol{\log \pi}=\text{MetropolisHastings}(q=q,\ \boldsymbol{\Sigma}=\boldsymbol{\Sigma},\ \boldsymbol{\theta}_0=\boldsymbol{\theta}_0,\ \hat n = \hat n$)
\end{algorithmic}
\end{algorithm}

If the potential scale reduction factor is $\hat{R}\ge 1.1$ for any of the estimated parameters (see~\cite{gelman2013bayesian}) and the chains are run for longer using the current iterations for the burnin phase. To allow for quicker convergence in the experimental datasets the matrix $\boldsymbol{\Sigma}$ scaled by a temperature factor $\alpha>0$ which controls parameter exploration.

\subsection*{S1.1 Python functions overview}

The file \texttt{functions\_MCMC.py} contains the general functions required to run the MCMC algorithm, as well as saving and plotting the results obtained.

The function \texttt{n\_state\_model(delta\_t, T, V, Lambda, P, sigma)} can be run to obtain the model simulation with the given parameters and returns the full information about the agent exact locations \texttt{x\_points}, noisy measured locations \texttt{y\_points}, times of switching \texttt{t}, as well as number of state switches \texttt{Nswitches} and states attained \texttt{states} for each measured interval. Hence, this function gives the full information on the hidden state evolution.

The function \texttt{get\_data\_dy(delta\_t, T, theta, get\_parameters, seed, correlated = True)} computes and returns the location increments \texttt{delta\_Y} obtained by calling the function \texttt{n\_state\_model} with the given settings, in which \texttt{get\_parameters} is a function of \texttt{theta} that needs to be specified that obtains the model parameters \texttt{V, Lambda, P, sigma} from the array of parameters to estimate \texttt{theta}. The \texttt{delta\_Y} obtained setting \texttt{correlated = True} are subsequent increments from a track, while setting \texttt{correlated = False} leads to a set of \texttt{N:=T/delta\_t} uncorrelated increments (or its integer part), obtained from \texttt{N} distinct datasets. The variable \texttt{seed} is used here, as well as in other functions, to obtain reproducible results and generate distinct datasets when needed.

The function \texttt{approx\_pdf\_up\_to\_1\_switch(V, Lambda, P, sigma, delta\_t, delta\_y)} computes the up-to-one-switch approximation for the probability of measuring a single location increment \texttt{delta\_y}. If \texttt{delta\_y} is a array of location increments, then it returns a array which entries correspond to the up-to-one-switch approximation PDF of measuring each increment in \texttt{delta\_y}. The function \texttt{approx\_pdf\_up\_to\_2\_switch(V, Lambda, P, sigma, delta\_t, delta\_y)} does the same but considers up-to-two state switches per measured interval. The function \texttt{approx\_pdf\_track\_up\_to\_1\_switch(V, Lambda, P, sigma, delta\_t, delta\_y)} returns a array which entries correspond to the up-to-one-switch approximation measuring each entry of \texttt{delta\_y}, conditioned on all the previous ones, which in this case is the set of correlated subsequent location increments obtained from a track. Details on how these approximations were obtained are provided in our previous work \cite{ceccarelli2025}. The function \texttt{approx\_pdf\_theta(theta, get\_parameters, delta\_t, delta\_y, up\_to\_switches = 1, track = False)} returns the results of one of the three approximations chosen by setting the parameter \texttt{up\_to\_switches} to 1 or 2, and for \texttt{up\_to\_switches = 1} by setting the boolean variable \texttt{track} to specify if the joint track PDF should be used.

The function \texttt{log\_pi\_hat(theta, get\_parameters, delta\_t, delta\_y, up\_to\_switches = 1, track = False)} gives the approximate log-likelihood of the parameter set \texttt{theta} given the data \texttt{delta\_y}, by summing the natural logarithm of the entries of the array returned by the function \texttt{approx\_pdf\_theta}. 

The function \texttt{MetropolisHastings(q, cov\_matrices, theta\_0, get\_parameters, burnin, n\_after\_burnin, n\_chains, delta\_t, delta\_Y, seed, up\_to\_switches = 1, track = False)} runs \texttt{burnin + n\_after\_burnin} Metropolis-Hastings MCMC iterations for a set number of chains (\texttt{n\_chains}) and returns \texttt{theta\_t} and \texttt{log\_pi} of the last \texttt{n\_after\_burnin} iterations. The data array used is specified in the variable \texttt{delta\_Y}, and the variables \texttt{up\_to\_switches = 1} and \texttt{track = False} allow to select the likelihood accessed through the globally defined \texttt{log\_pi\_hat}. The variable \texttt{theta\_0} specifies the starting values of the Markov chain and it is a array of shape \texttt{(n\_param, n\_chains)}, where \texttt{n\_param} is the number of parameters to estimate. The proposed parameters of the Markov chain are sampled using the proposal density \texttt{q}, which we define as multivariate normal distribution \texttt{q(theta, cov\_matrix)} with mean the current parameter set \texttt{theta} and by covariance the matrix \texttt{cov\_matrix} of shape \texttt{(n\_param, n\_param)} selected to be \texttt{cov\_matrix = cov\_matrices[ind,:,:]} where \texttt{ind} is the chain index and \texttt{cov\_matrices} is a array of shape \texttt{(n\_chains, n\_param, n\_param)}. We note that in this \texttt{q} is symmetric and this allows us to reduce the algorithm. 

We also write down the function \texttt{distinct\_track\_runs\_MCMC} that allows us to run the MCMC separately for a number distinct datasets. Each dataset contains a set of location increments usually generated by a distinct track.

\section*{S2 Supplementary results}\label{SI:S2}

\subsection*{S2.1 Calibration of a two-state model and comparison between the track likelihood and the marginal likelihood}\label{SI:S2.1}

Here, we discuss the use of marginal likelihood $P_1^{\mathcal{M}}$, which is an alternative to the track likelihood $P_1$ (used in Figure~2). For the calibration of the two-state model, the track likelihood and the marginal likelihood produce very similar posterior distributions (Supplementary Information Figure~S2\textbf{A} and S3) and a similar likelihood approximation for a single location increment (Supplementary Information Figure~S2\textbf{B}). The only exception is the positive correlation obtained using the marginal likelihood in the bivariate posterior of $\log(\lambda_1)$ and $\log(\lambda_2)$ which is not present when using the track likelihood.

Further observations can be made to assess for which parameter regimes the likelihood approximations are valid. In particular, state switches need to occur sufficiently infrequently compared to the data collection frequency. In other words, the switching rates $\lambda$ need to be sufficiently small compared to $\Delta t$. Indeed, the averages of the parameters in the parameter sets giving the maximum likelihood, obtained calibrating the model to 100 distinct datasets across different parameter regimes, are valid estimates for the true parameters within their standard deviation for small switching rates, $\lambda_1=2\lambda_2=1,2,4$ (Supplementary Information Figure~S4\textbf{A}-\textbf{E}). On the other hand, for $\lambda_1=2\lambda_2=8$ using the track likelihood leads to closer estimates for the model parameters to their true value than using the marginal likelihood. In contrast, for high rates ($\lambda_1=2\lambda_2=16$) the true model parameters cannot be recovered using both the marginal and the track likelihood. This behaviour is expected since the likelihood approximation is less accurate as the rates increase (for example Supplementary Information Figure~S4\textbf{F}-\textbf{G}). In particular, from Supplementary Information Figure~S4\textbf{F}-\textbf{G}, we note that the peaks present in the single-location-increment distributions are not clearly distinguishable for high rates. Hence, plotting the data distributions can give a first indication as to whether the data collection frequency is sufficient to capture the underlying process.

The standard deviation of the posteriors, which represents the uncertainty in the estimated parameters, can be reduced by increasing the track length, both using the track likelihood and the marginal likelihood. In particular, for both likelihoods the median standard deviations decrease as the number of data points $N$ is increased and, specifically, the median standard deviation and $1/\sqrt{N}$ are linearly dependent, as expected (Supplementary Information Figure~S5\textbf{A}-\textbf{E}).

Finally, we point out that the main advantage of using the marginal likelihood over the track likelihood is the significantly lower runtime. In particular, the runtime of the four MCMC chains for the marginal likelihood is between 70 and 100 seconds for the model in Figure~2\textbf{B}, and is not significantly varied with the number of data points $N=50, 100, 200$ (the red line in Supplementary Information Figure~S5\textbf{F}). In contrast, the runtime for the track likelihood is approximately proportional to the number of data points used, and is approximately 0.6, 1.2, 2.4 hours for datasets of $N=50, 100, 200$ data points, respectively (the blue line in Supplementary Information Figure~S5\textbf{F}).

\begin{figure*}[!ht]
    \centering
    \includegraphics[width=1\textwidth]{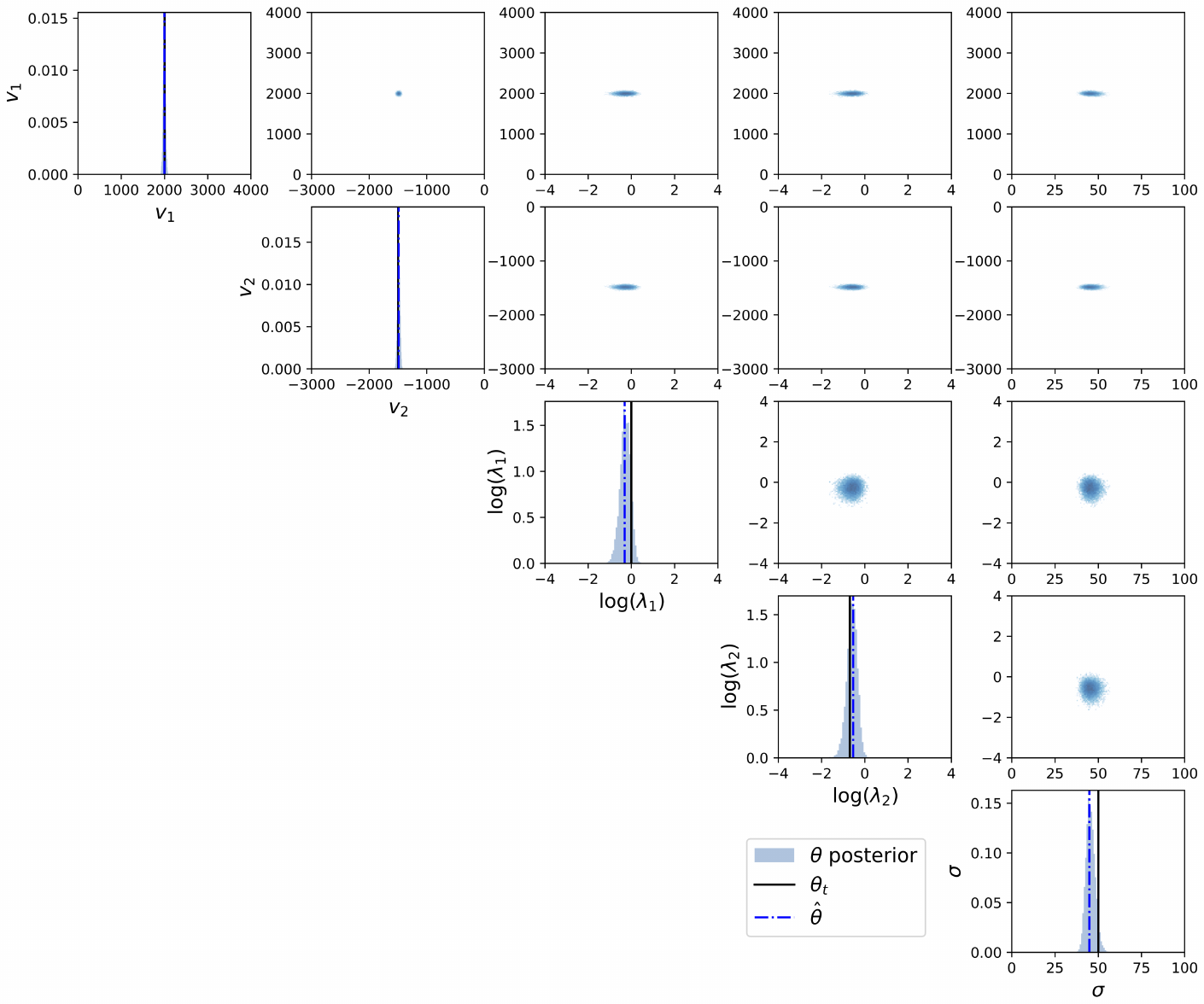}
    \caption{\textbf{Comparison of the univariate and bivariate posterior distributions with the priors corresponding to Figure~2}. The chains are run using the track likelihood $P_1$, with burnin 10000 followed by 10000 MCMC iterations. Posterior distributions are plotted with horizontal-axis scale corresponding to the uniform prior used to sample the initial parameter sets $[0,4000]\times [-3000,0]\times[-4,4]\times[-4,4]\times[0,100]$. The proposals are Gaussian distributions with adaptive covariance matrix. The initial covariance matrix is set to the diagonal matrix with diagonal $[0.1, 0.1, 0.0001, 0.0001, 0.01]$.}
    \label{SI_Fig_S1}
\end{figure*}

\begin{figure*}[!ht]
    \centering
    \includegraphics[width=1\textwidth]{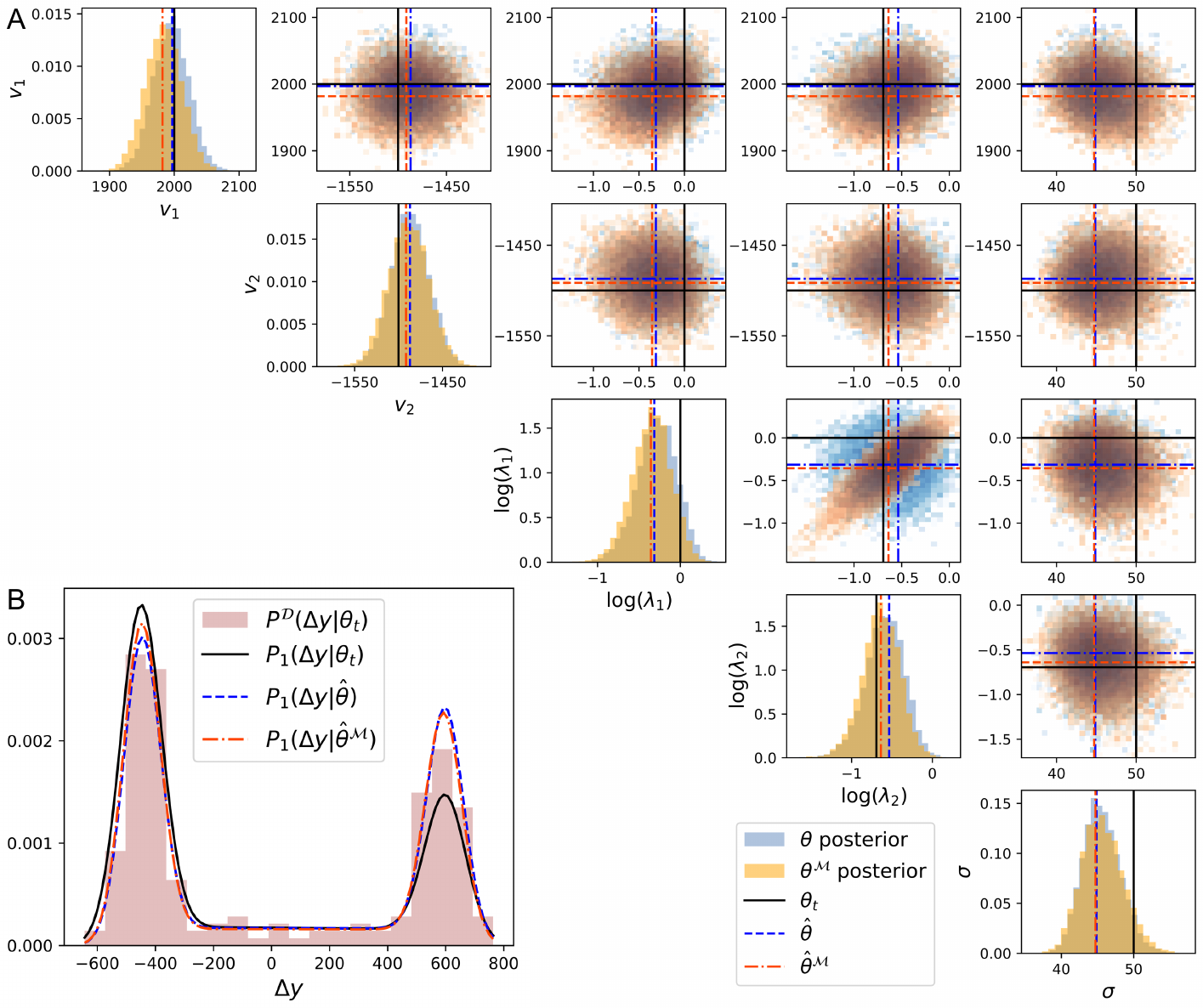}
    \caption{\textbf{Comparison of results obtained using the track likelihood $P_1$ and the marginal likelihood $P_1^\mathcal{M}$ to calibrate the two-state model in Figure~2\textbf{B}}. \textbf{A} shows the univariate and bivariate posterior distributions using the track likelihood (blue) and using the marginal likelihood (orange) for the data shown in Figure~2\textbf{C}. \textbf{B} compares the approximate PDF for a single increment $P_1(\Delta y)$ for the parameter set that maximises the track likelihood, $\boldsymbol{\hat \theta}$, and for the parameter set that maximises the marginal likelihood, $\boldsymbol{\hat \theta}^{\mathcal{M}}$, to the approximate PDF at the true parameter set $\boldsymbol{\theta}_t$ and to the data distribution $P^{\mathcal{D}}$. Uniform priors are used to sample the initial parameter sets in the intervals $[0,4000]\times [-3000,0]\times[-4,4]\times[-4,4]\times[0,100]$. The proposals are Gaussian distributions with adaptive covariance matrix, and burnin is set to 10000 followed by 10000 MCMC iterations.}
    \label{SI_Fig_S2}
\end{figure*}

\begin{figure*}[!ht]
    \centering
    \includegraphics[width=1\textwidth]{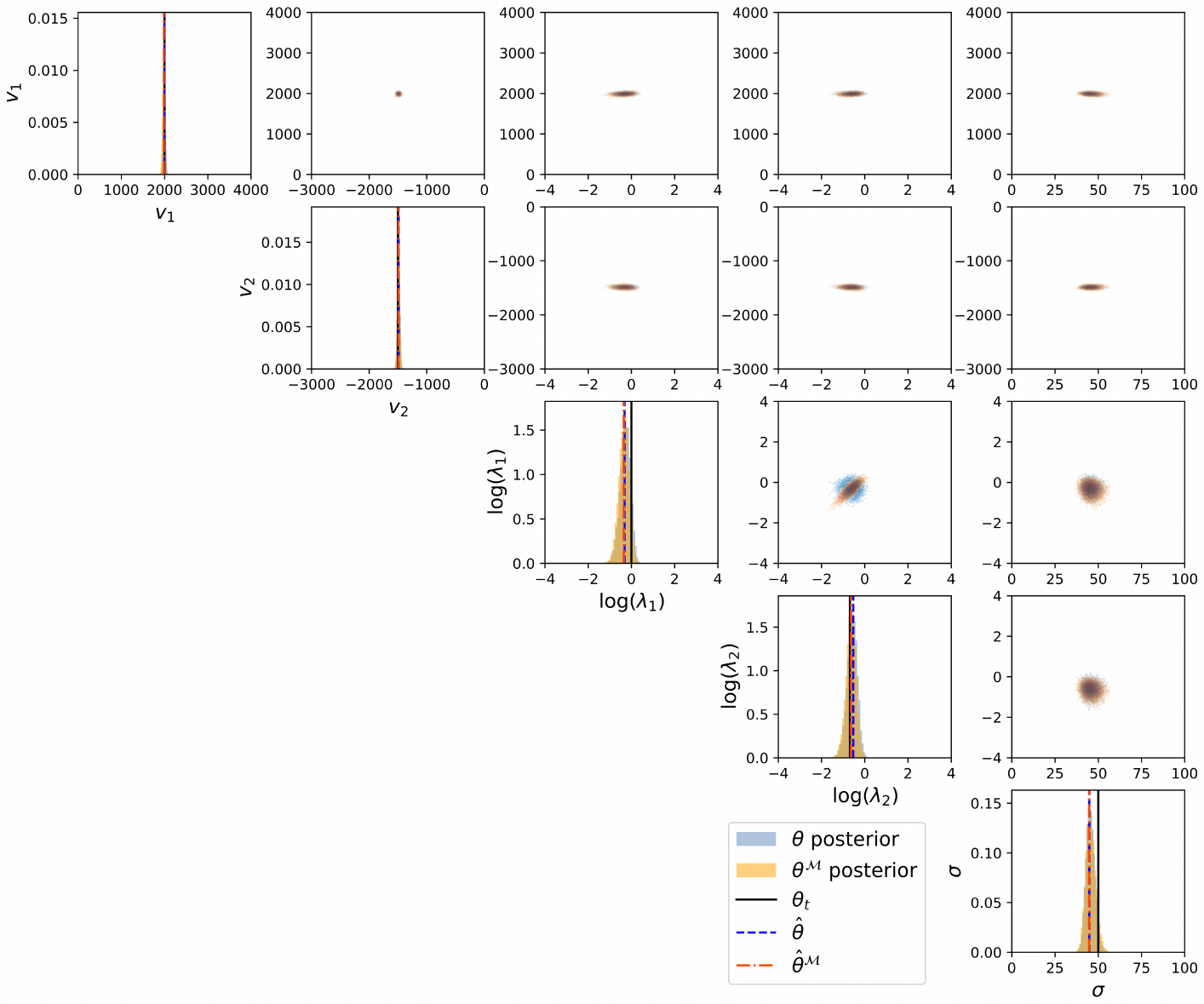}
    \caption{\textbf{Comparison of the  univariate and bivariate posterior distributions with the priors used to calibrate the two-state model in Figure~2\textbf{B}, corresponding to the results in Figure~S2}. The posterior distributions are plotted  with horizontal-axis corresponding to the uniform priors used, in the intervals $[0,4000]\times [-3000,0]\times[-4,4]\times[-4,4]\times[0,100]$.}
    \label{SI_Fig_S3}
\end{figure*}

\begin{figure*}[!ht]
    \centering
    \includegraphics[width=0.9\textwidth]{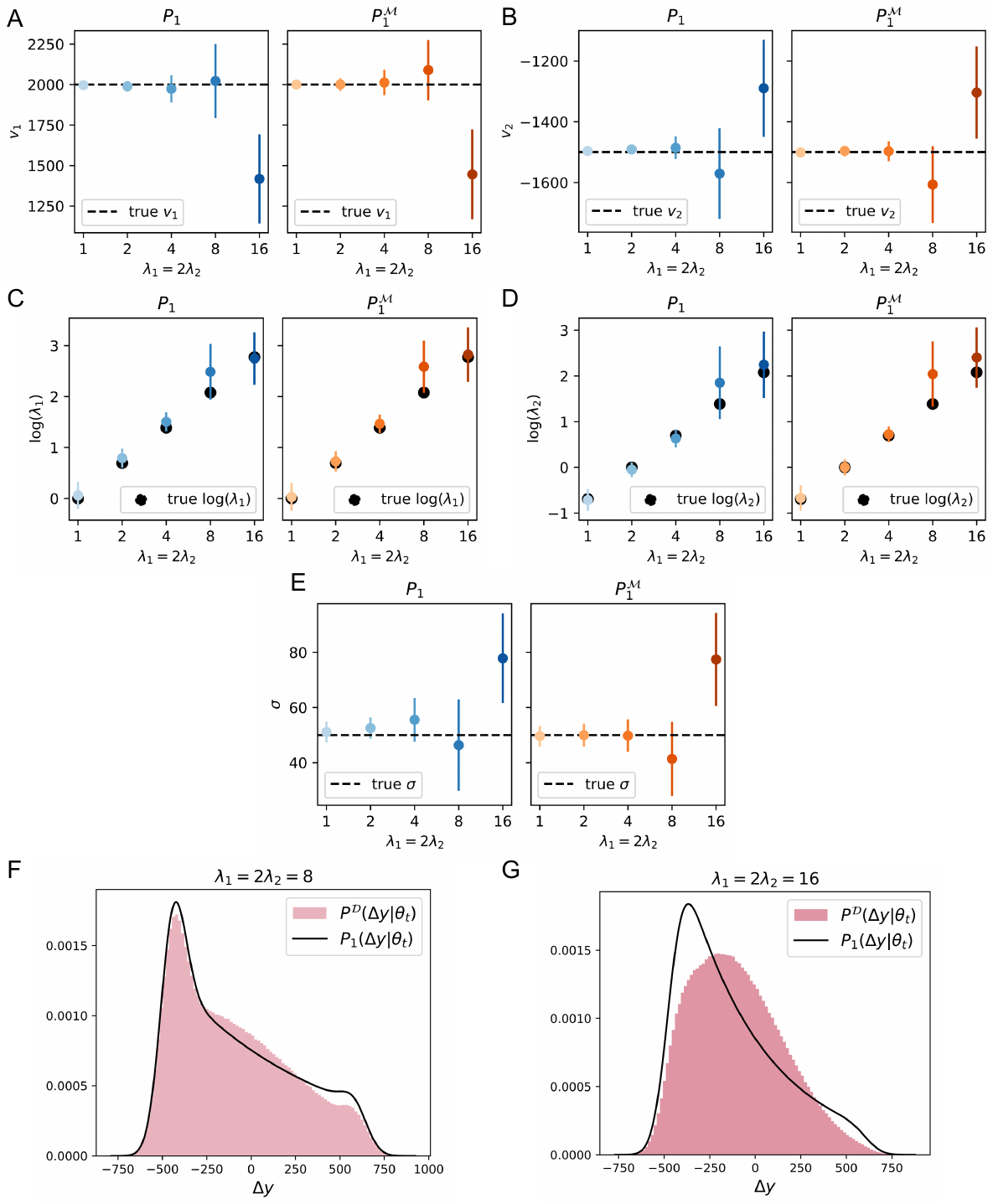}
    \caption{\textbf{Comparison of estimated parameter sets that give the maximum likelihood across 100 datasets for the two-state model in Figure~2\textbf{B}}. \textbf{A}-\textbf{E} show the means and standard deviations of the estimated parameters in the set that maximises likelihood across 100 datasets, as the rates are increased ($\lambda_1=2\lambda_2$ are varied in the set $\{1,2,4,8,16\}$). In each panel the left plot is obtained using the track likelihood $P_1$, while the right plot is obtained using the marginal likelihood $P_1^\mathcal{M}$. Uniform priors are used to sample the initial parameter sets in the intervals $[0,4000]\times [-3000,0]\times[-4,4]\times[-4,4]\times[0,100]$. The proposals are Gaussian distributions with adaptive covariance matrix, and burnin is set to 10000 followed by 10000 MCMC iterations. \textbf{F} and \textbf{G} compare the empirical PDF for a single increment $P^{\mathcal{D}}$ with the approximate PDF $P_1$ for parameter sets with high switching rates, $\lambda_1=2\lambda_2=8$ and $\lambda_1=2\lambda_2=16$, respectively.}
    \label{SI_Fig_S4}
\end{figure*}

\begin{figure*}[!ht]
    \centering
    \includegraphics[width=1\textwidth]{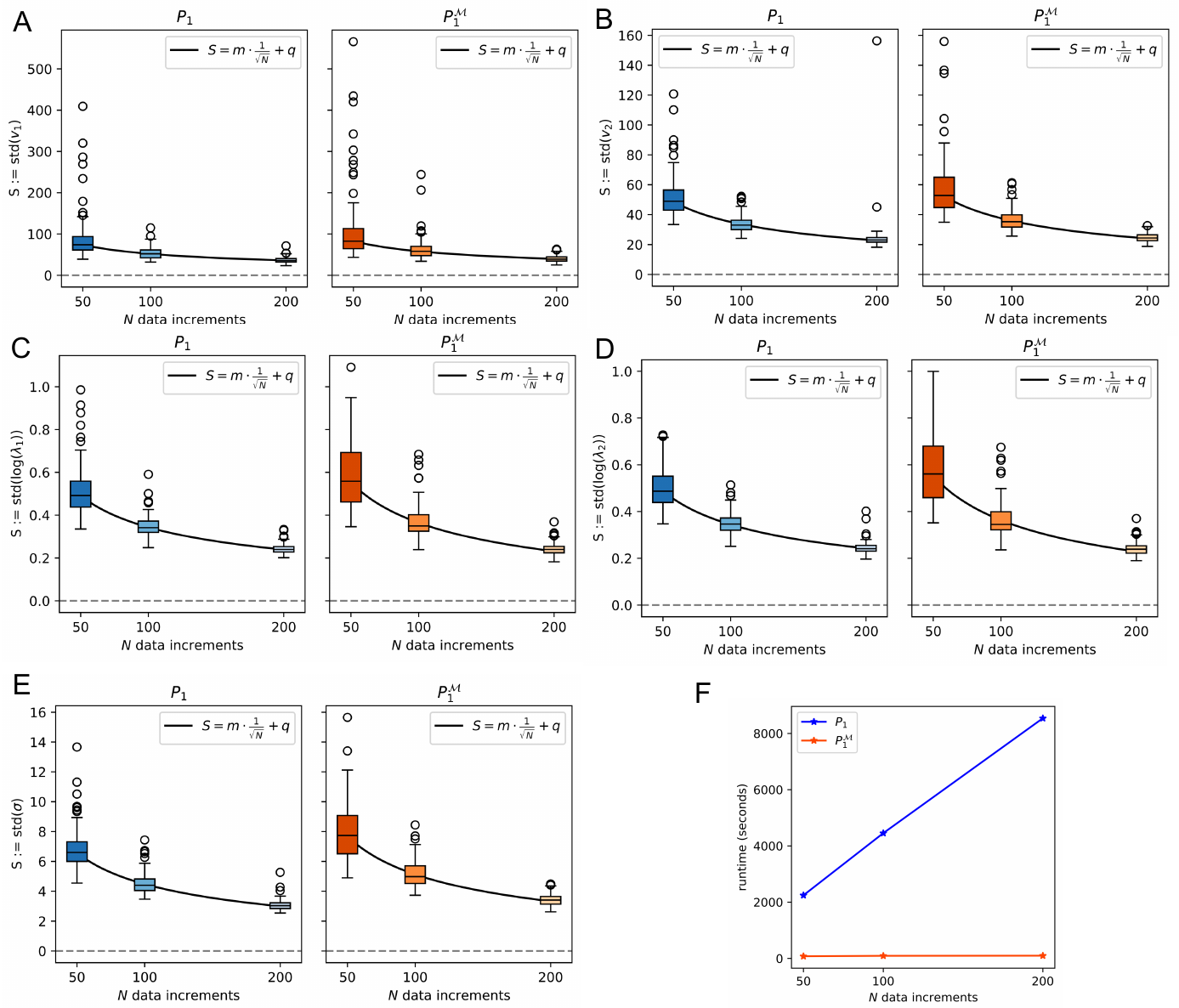}
    \caption{\textbf{Comparison of the standard deviations of the estimated parameter posteriors obtained across 100 datasets for the two-state model in Figure~2\textbf{B}}. The standard deviations of the 100 posteriors obtained calibrating 100 datasets is shown in \textbf{A}-\textbf{E}, as the number of data points are varied, $N=50$, $N=100$ and $N=200$. In each panel the left plot is obtained using the track likelihood, while the right plot is obtained using the marginal likelihood. \textbf{F} compares the runtime required for a single simulation with burnin 10000, followed by 10000 MCMC iterations. The runtime for the marginal likelihood is constant (around 80-85 seconds), while it increases proportionally with the number of data points for the track likelihood (0.6, 1.2, 2.4 hours for datasets of $N=50, 100, 200$ data points, respectively).}
    \label{SI_Fig_S5}
\end{figure*}

\subsection*{S2.2 Other Supplementary Information figures}

Here, we show additional figures.

\begin{figure*}[!ht]
    \centering
    \includegraphics[width=1\textwidth]{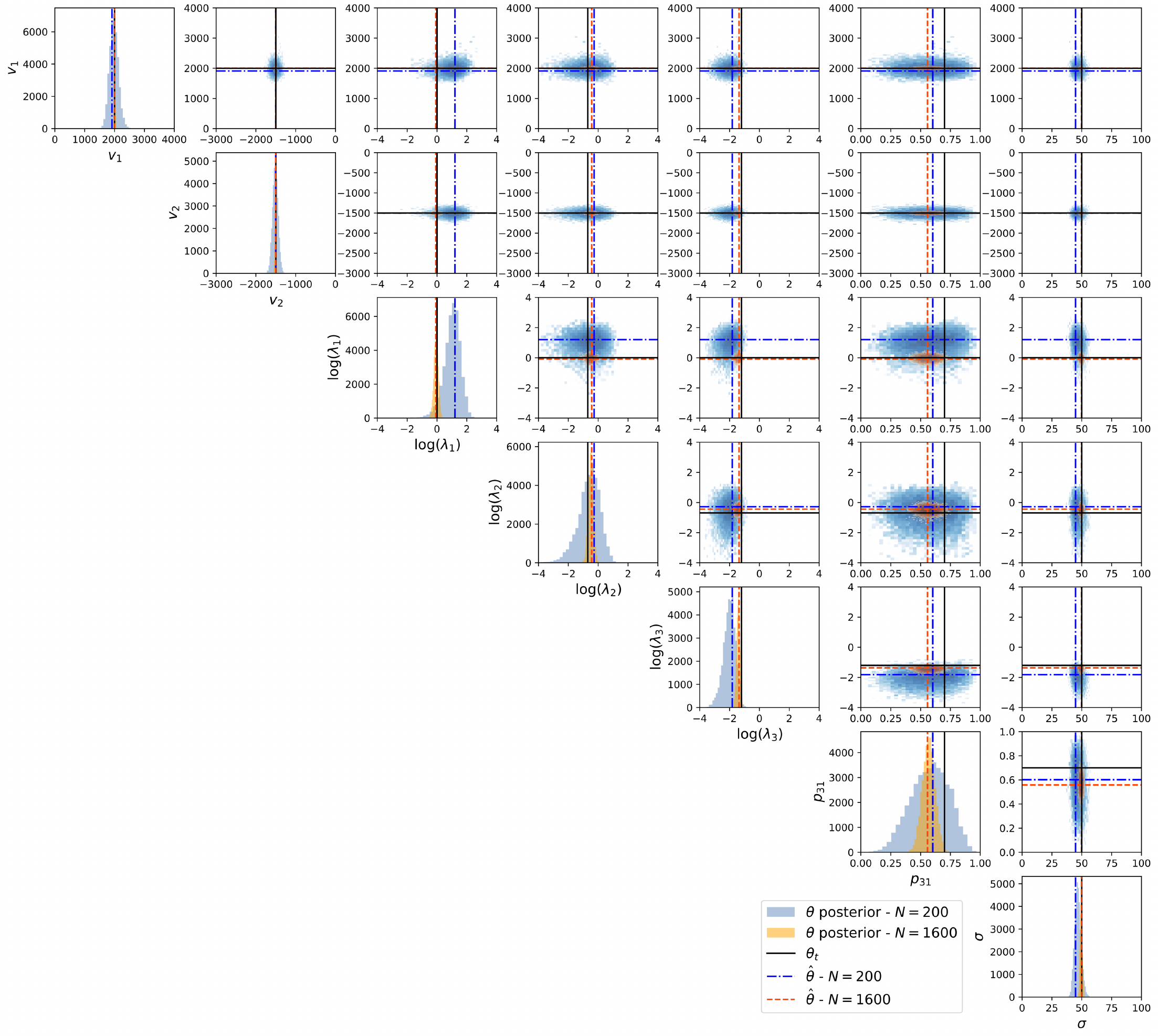}
    \caption{\textbf{Comparison of the  univariate and bivariate posterior distributions with priors used to calibrate the three-state model in Figure~3\textbf{B} using track likelihood and marginal likelihood}. The posterior distributions are plotted with horizontal axes corresponding to the full prior used $[0,4000]\times [-3000,0]\times[-4,4]\times[-4,4]\times[-4,4]\times[0,1]\times[0,100]$ for the data shown in Figure~3\textbf{B}, the use of track likelihood and marginal likelihood, reproducing the results in Figure~S2.}
    \label{SI_Fig_S6}
\end{figure*}

\begin{figure*}[!ht]
    \centering
    \includegraphics[width=1\textwidth]{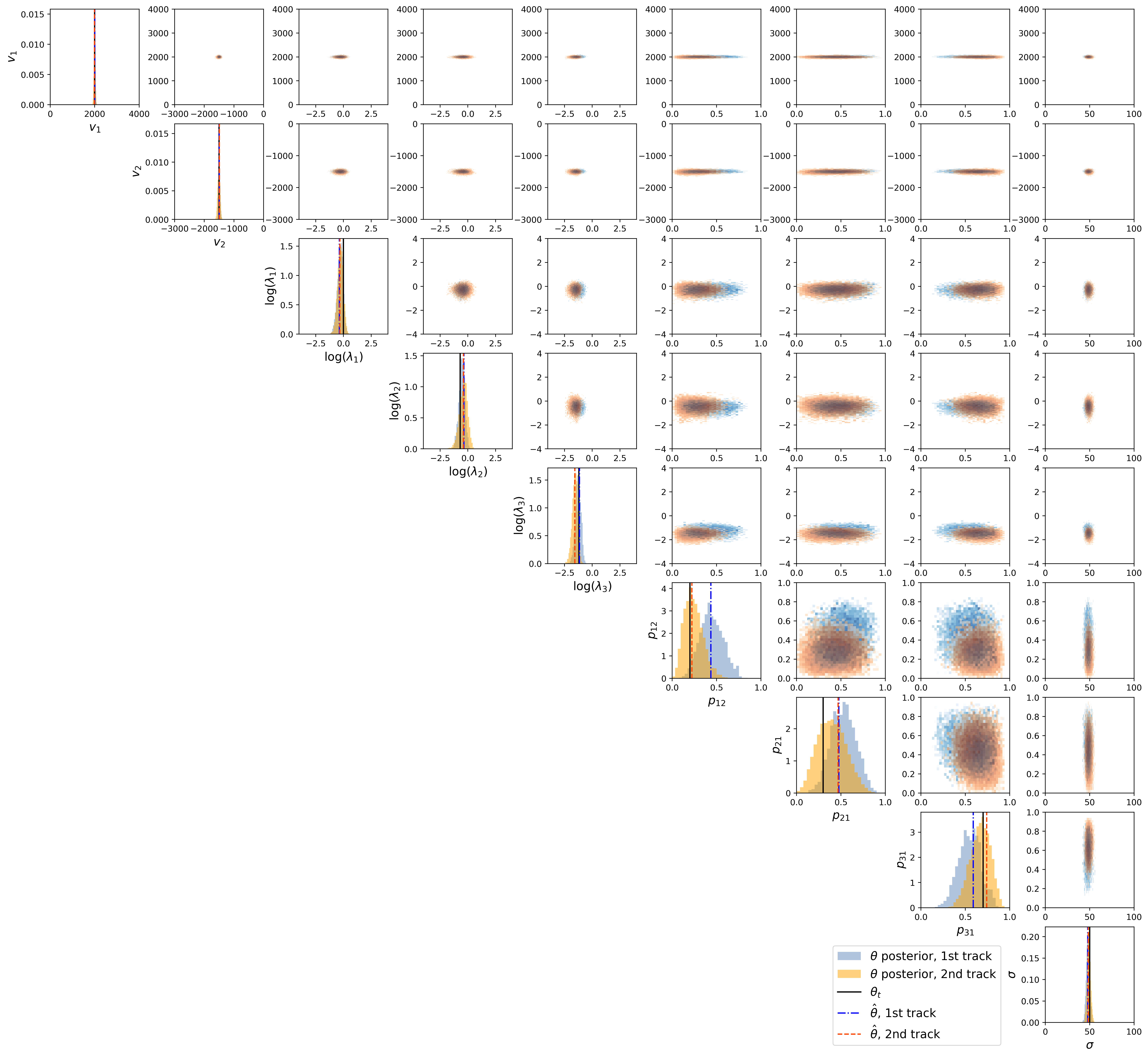}
    \caption{\textbf{Comparison of the  univariate and bivariate posterior distributions for all estimated parameters with priors used to calibrate the three-state model in Figure~4\textbf{E} for the 1st and 2nd track}. The posterior distributions are plotted with horizontal axes corresponding to the priors used $[0,4000]\times [-3000,0]\times[-4,4]\times[-4,4]\times[-4,4]\times[0,1]\times[0,1]\times[0,1]\times[0,100]$ for the data shown in Figure~4\textbf{C} (left). The proposals are Gaussian distributions with adaptive covariance matrix. The burnin was set to 10000 iterations, followed by 10000 MCMC iterations.}
    \label{SI_Fig_S7}
\end{figure*}

\begin{figure*}[!ht]
    \centering
    \includegraphics[width=1\textwidth]{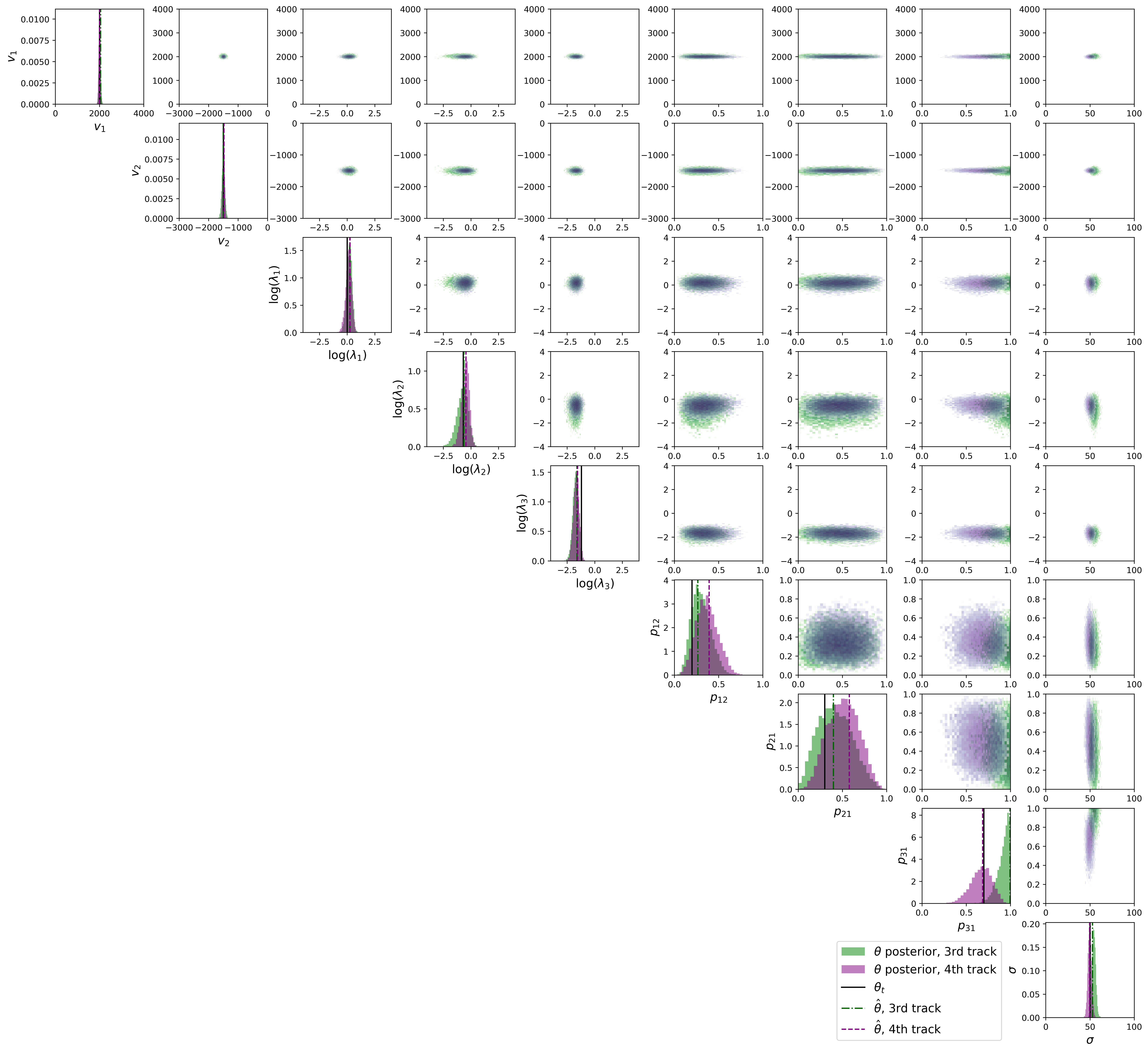}
    \caption{\textbf{Comparison of the  univariate and bivariate posterior distributions for all estimated parameters with priors used to calibrate the three-state model in Figure~4\textbf{E} for the 3rd and 4th track}. The posterior distributions are plotted with horizontal axes corresponding to the priors used $[0,4000]\times [-3000,0]\times[-4,4]\times[-4,4]\times[-4,4]\times[0,1]\times[0,1]\times[0,1]\times[0,100]$ for the data shown in Figure~4\textbf{C} (right).}
    \label{SI_Fig_S8}
\end{figure*}

\begin{figure*}[!ht]
    \centering
    \includegraphics[width=1\textwidth]{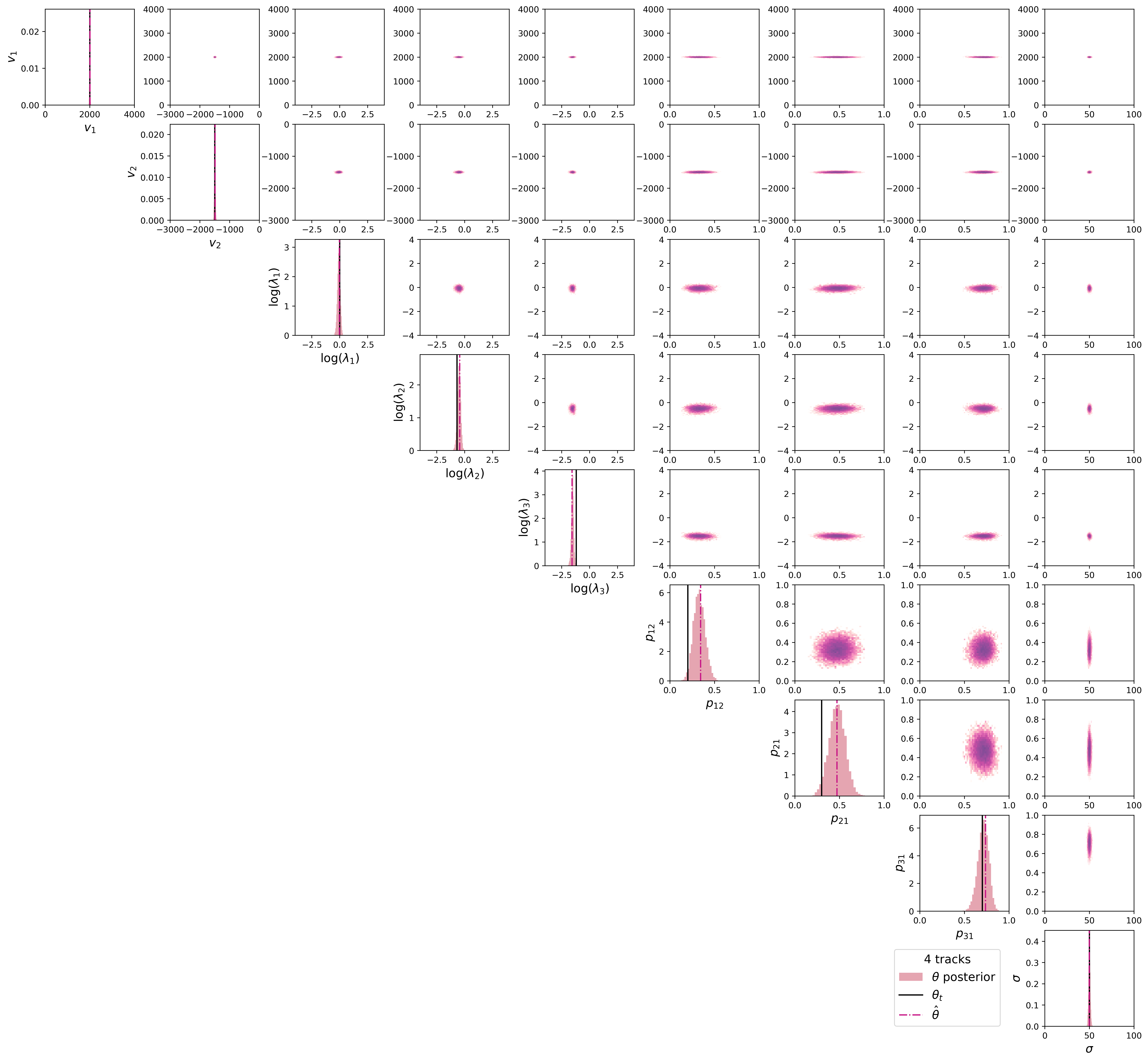}
    \caption{\textbf{Comparison of the univariate and bivariate posterior distributions for all estimated parameters with priors used to calibrate the three-state model in Figure~4\textbf{E} for four tracks used together to compute the likelihood}. The posterior distributions are plotted with horizontal axes corresponding to the priors used $[0,4000]\times [-3000,0]\times[-4,4]\times[-4,4]\times[-4,4]\times[0,1]\times[0,1]\times[0,1]\times[0,100]$ for the four data tracks shown in Figure~4\textbf{C}.}
    \label{SI_Fig_S9}
\end{figure*}

\begin{figure*}[!ht]
    \centering
    \includegraphics[width=1\textwidth]{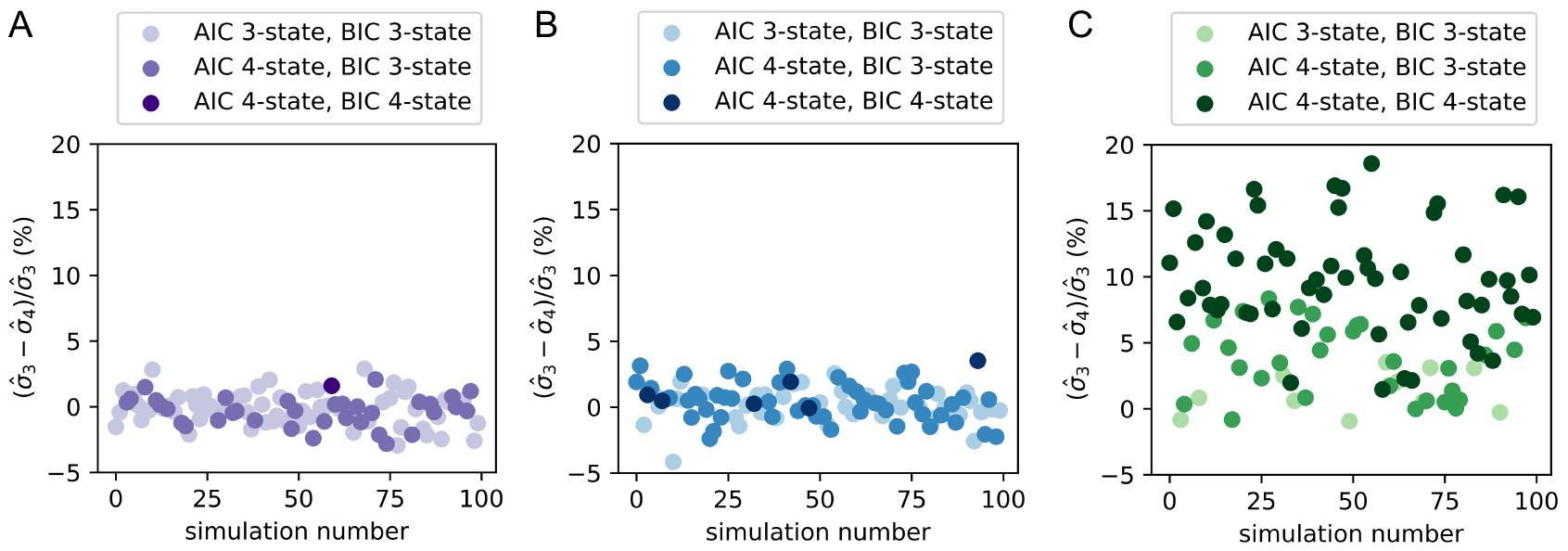}
    \caption{\textbf{Relative difference in the noise standard deviation in percentage estimated calibrating a three-state model versus a four-state model using data generated using three distinct four-state models in Figure~6}. \textbf{A}-\textbf{C} show the relative difference in estimated noise standard deviation $(\hat\sigma_3-\hat\sigma_4)/\hat\sigma_3$ obtained calibrating three-state and four-state models to the 100 datasets generated from the models in Figure~6\textbf{A}-\textbf{C}, respectively.}
    \label{SI_Fig_S10}
\end{figure*}

\begin{figure*}[!ht]
    \centering
    \includegraphics[width=1\textwidth]{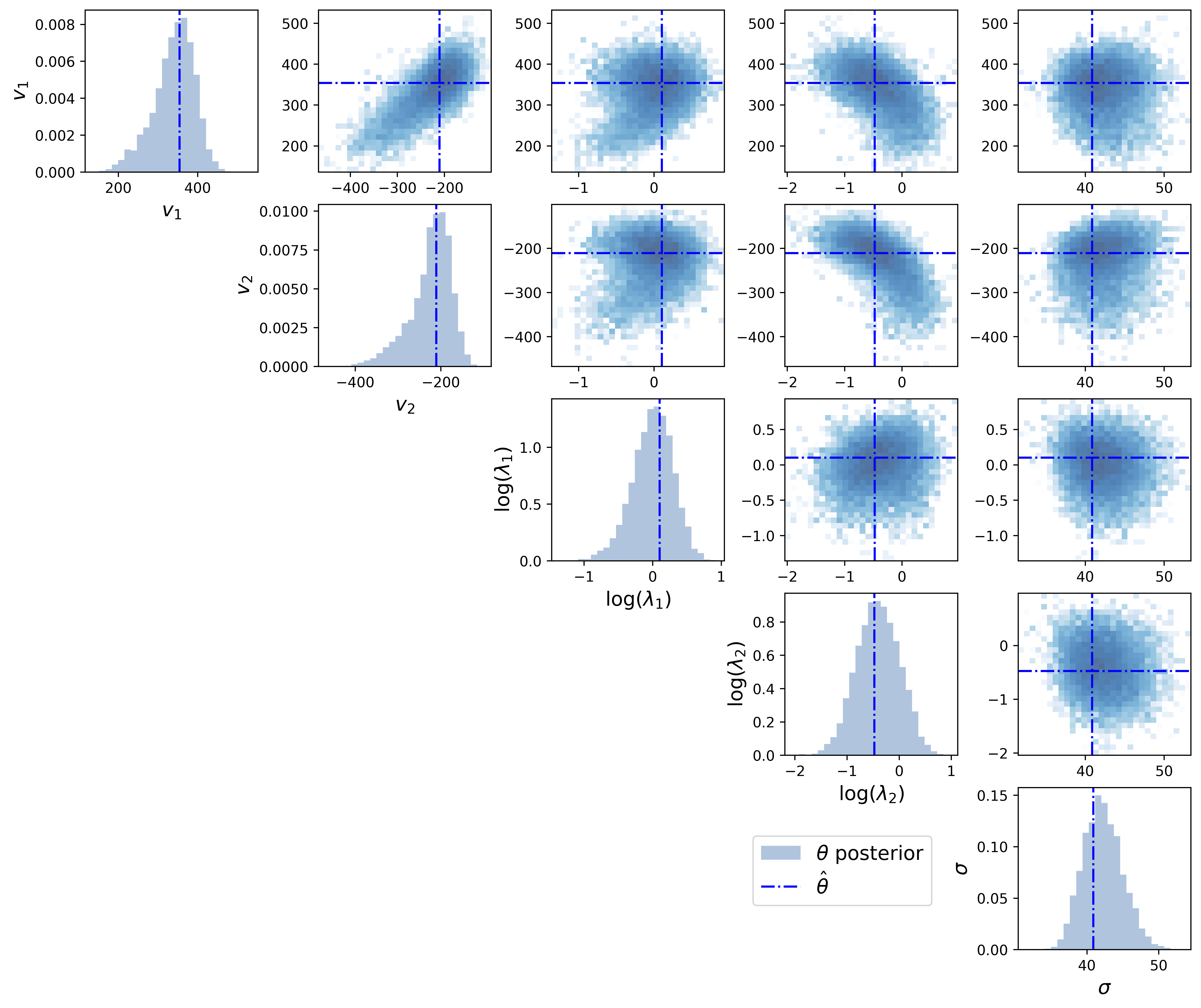}
    \caption{\textbf{Posteriors obtained from the calibration of a two-state model for track A in Figure~8A}. The prior is uniform in $[0, 2*\max\{\Delta y_1,\ldots,\Delta y_N\}/\Delta t]\times [2*\min\{\Delta y_1,\ldots,\Delta y_N\}/\Delta t, 0]\times [-4,4]\times[-4,4]\times[0,100]$. The temperature is set to $\alpha=1$.}
    \label{SI_Fig_S11}
\end{figure*}

\begin{figure*}[!ht]
    \centering
    \includegraphics[width=1\textwidth]{FIGURES/track_A_3-state_best_parameters_posteriors.png}
    \caption{\textbf{Posteriors obtained from the calibration of a two-state model for track A in Figure~8A}. The prior is uniform in $[0, 2*\max\{\Delta y_1,\ldots,\Delta y_N\}/\Delta t]\times [2*\min\{\Delta y_1,\ldots,\Delta y_N\}/\Delta t, 0]\times [-4,4]\times[-4,4]\times[-4,4]\times[0,1]\times[0,1]\times[0,1]\times[0,100]$. The temperature is set to $\alpha=1$.}
    \label{SI_Fig_S12}
\end{figure*}

\begin{figure*}[!ht]
    \centering
    \includegraphics[width=1\textwidth]{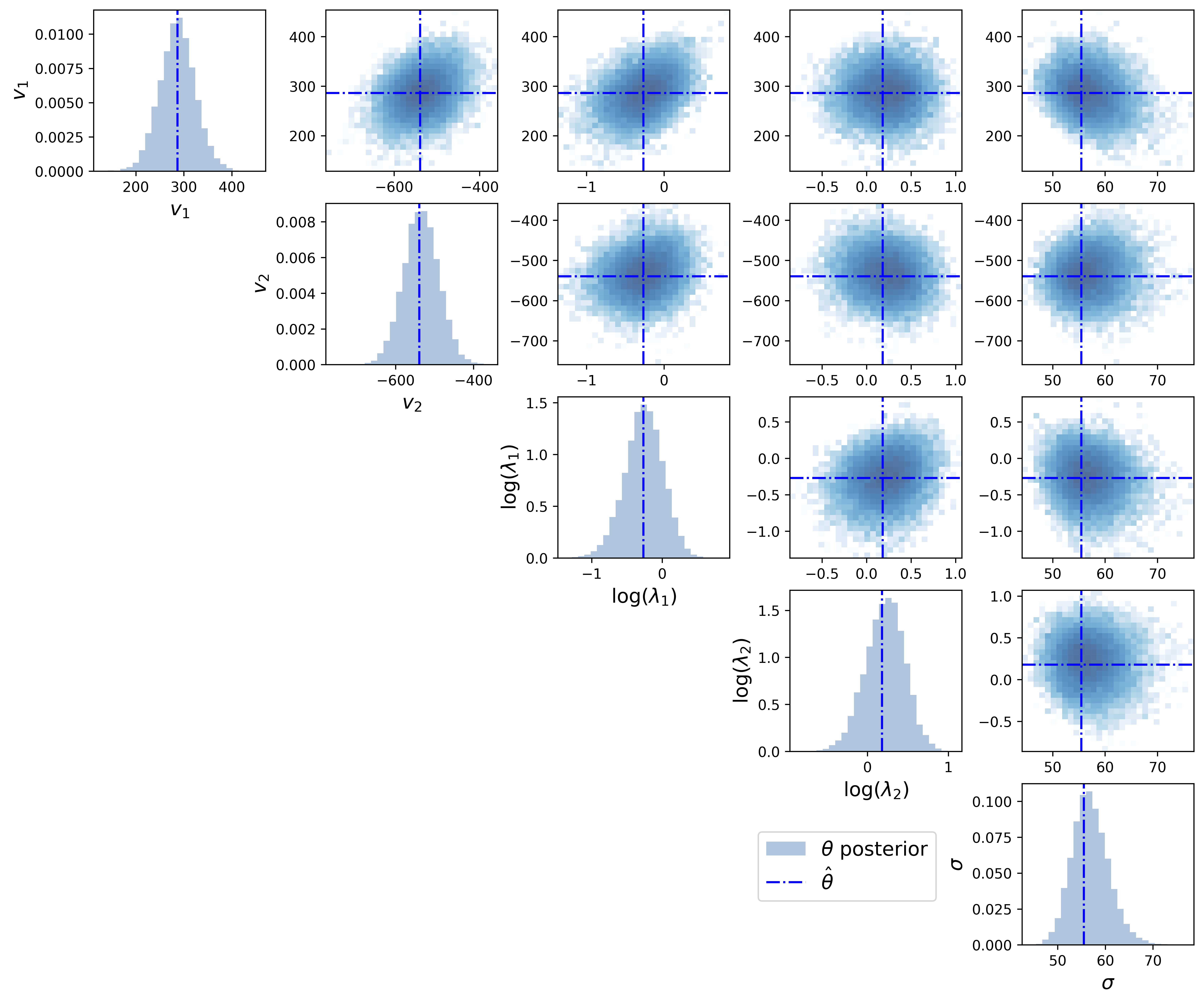}
    \caption{\textbf{Posteriors obtained from the calibration of a two-state model for track B in Figure~8A}. The prior is uniform in $[0, 2*\max\{\Delta y_1,\ldots,\Delta y_N\}/\Delta t]\times [2*\min\{\Delta y_1,\ldots,\Delta y_N\}/\Delta t, 0]\times [-4,4]\times[-4,4]\times[0,100]$. The temperature is set to $\alpha=0.3$.}
    \label{SI_Fig_S13}
\end{figure*}

\begin{figure*}[!ht]
    \centering
    \includegraphics[width=1\textwidth]{FIGURES/track_B_3-state_best_parameters_posteriors.png}
    \caption{\textbf{Posteriors obtained from the calibration of a two-state model for track B in Figure~8A}. The prior is uniform in $[0, 2*\max\{\Delta y_1,\ldots,\Delta y_N\}/\Delta t]\times [2*\min\{\Delta y_1,\ldots,\Delta y_N\}/\Delta t, 0]\times [-4,4]\times[-4,4]\times[-4,4]\times[0,1]\times[0,1]\times[0,1]\times[0,100]$. The temperature is set to $\alpha=0.3$.}
    \label{SI_Fig_S14}
\end{figure*}

\begin{figure*}[!ht]
    \centering
    \includegraphics[width=1\textwidth]{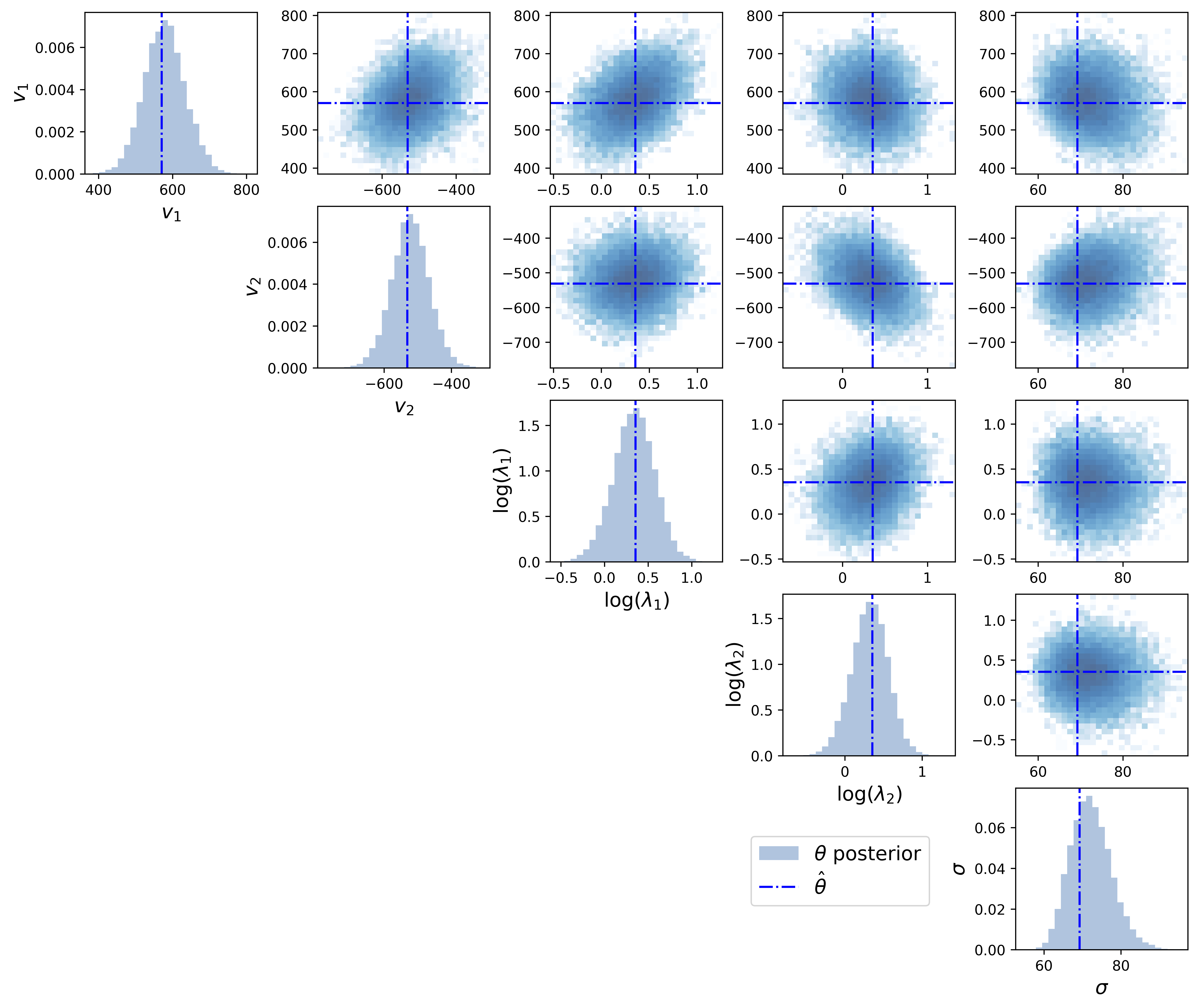}
    \caption{\textbf{Posteriors obtained from the calibration of a two-state model for track C in Figure~8A}. The prior is uniform in $[0, 2*\max\{\Delta y_1,\ldots,\Delta y_N\}/\Delta t]\times [2*\min\{\Delta y_1,\ldots,\Delta y_N\}/\Delta t, 0]\times [-4,4]\times[-4,4]\times[0,100]$. The temperature is set to $\alpha=0.3$.}
    \label{SI_Fig_S15}
\end{figure*}

\begin{figure*}[!ht]
    \centering
    \includegraphics[width=1\textwidth]{FIGURES/track_C_3-state_best_parameters_posteriors.png}
    \caption{\textbf{Posteriors obtained from the calibration of a two-state model for track C in Figure~8A}. The prior is uniform in $[0, 2*\max\{\Delta y_1,\ldots,\Delta y_N\}/\Delta t]\times [2*\min\{\Delta y_1,\ldots,\Delta y_N\}/\Delta t, 0]\times [-4,4]\times[-4,4]\times[-4,4]\times[0,1]\times[0,1]\times[0,1]\times[0,100]$. The temperature is set to $\alpha=0.3$.}
    \label{SI_Fig_S16}
\end{figure*}

\begin{figure*}[!ht]
    \centering
    \includegraphics[width=1\textwidth]{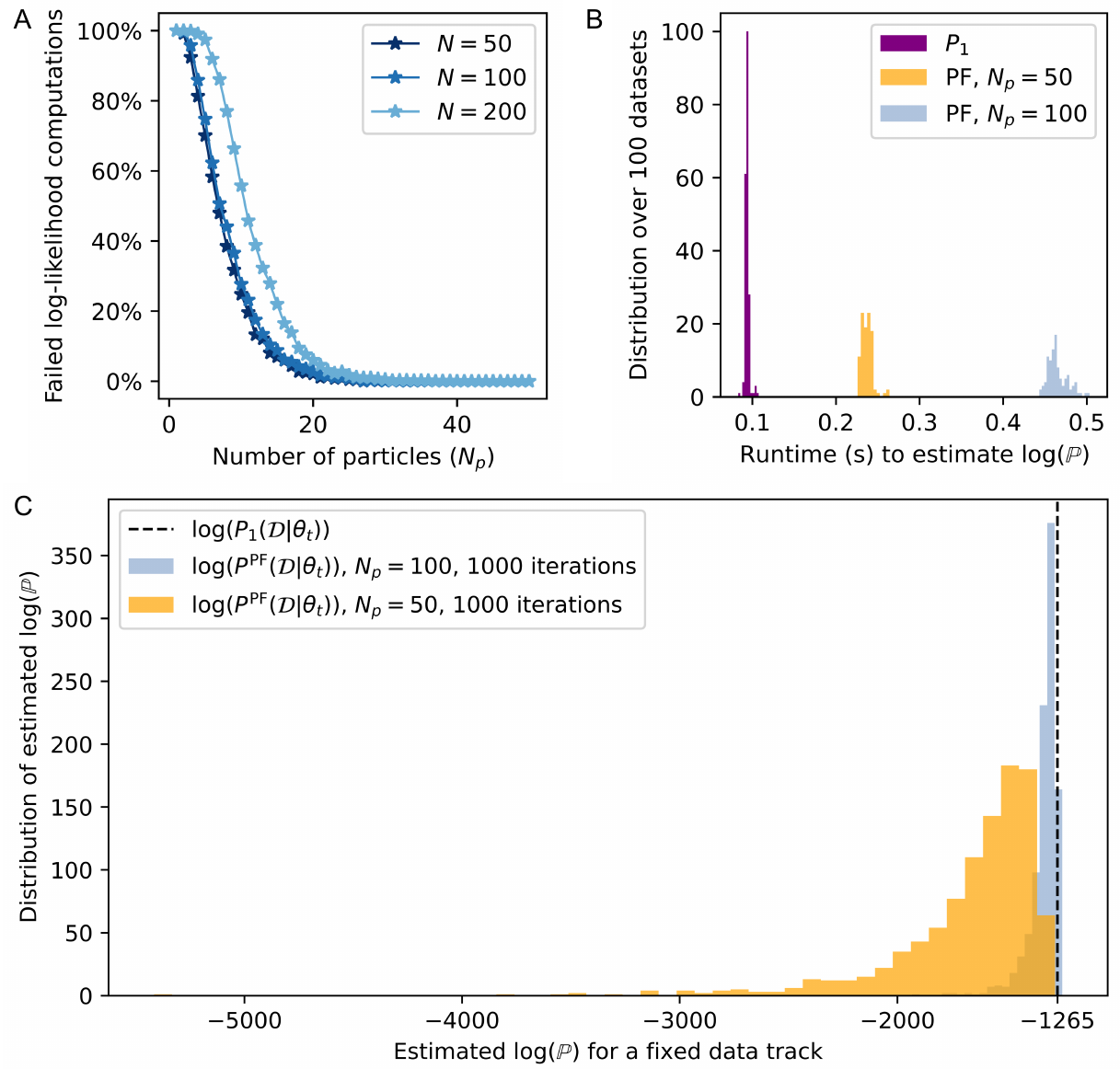}
    \caption{\textbf{Comparison of the framework based on the approximate track likelihood $P_1$ with particle filtering (PF) methods}. The results shown are produced using the model in Figure~2\textbf{B}. \textbf{A} shows the percentage of degenerate log-likelihood computations as the number of particles $N_p$ in the particle filtering method varies, computed across 1000 simulations, for a number of data points $N=50, 100, 200$. \textbf{B} shows the runtime required to estimate the log-likelihood using the track PDF approximation $P_1$. \textbf{C} shows the value of the log-likelihood estimated using $P_1$, and using the particle filtering with $N_p=50$ and $N_p=100$, computed across 1000 simulations.}
    \label{SI_Fig_S17}
\end{figure*}

\bibliography{sn-bibliography}